\documentclass[aps,pre,reprint,groupedaddress,showpacs]{revtex4-1}

\usepackage{graphicx}
\usepackage{subfigure}
\usepackage{ragged2e}
\usepackage{subfigure}
\usepackage{amsmath}
\usepackage{color} 

\newcommand{\one}{($i$) }
\newcommand{\two}{($ii$) }
\newcommand{\three}{($iii$) }

\newcommand{\onel}{($A$)}
\newcommand{\twol}{($B$)}

\newcommand{\sm}{\emph{Appendix}}

\newcommand{\fig}{FIG.}
\newcommand{\tab}{TABLE}

\begin{document}


\title{Facilitated diffusion buffers noise in gene expression}


\author{Armin P. Schoech}

\author{Nicolae Radu Zabet}
\email[Corresponding author: ]{n.r.zabet@gen.cam.ac.uk}
\affiliation{Cambridge Systems Biology Centre, University of Cambridge, Tennis Court Road, Cambridge CB2 1QR, UK}
\affiliation{Department of Genetics, University of Cambridge, Downing Street, Cambridge CB2 3EH, UK}



\begin{abstract}
Transcription factors perform facilitated diffusion (3D diffusion in the cytosol and 1D diffusion on the DNA) when binding to their target sites to regulate gene expression. Here, we investigated the influence of this binding mechanism on the noise in gene expression. Our results showed that, for biologically relevant parameters, the binding process can be represented by a two-state Markov model and that the accelerated target finding due to facilitated diffusion leads to a reduction in both the mRNA and the protein noise.
\end{abstract}


\maketitle

\section{Introduction}

Cellular reactions are fundamentally stochastic processes. Recent advances in single cell measurements have given insight into the details of some cellular processes and provided precise quantitative measurements of individual reactions \citep{golding_2005, elf_2007, hammar_2012}. This has allowed increasingly detailed modelling of cellular dynamics and a better understanding of the stochasticity of cellular processes. In particular, two fields have strongly benefited from this development: \one stochastic gene expression models (e.g. \citep{paulsson_2004,friedman_2006}) and \two models of transcription factor (TF) dynamics (e.g. \citep{mirny_2009,benichou_2011,zabet_2012_model}). Except for a few studies (e.g. \citep{zabet_2013_time,meyer_2012,pulkkinen_2013,sharon_2014}), the combined effects of these two were not investigated, despite the fact that they directly affect one another. 


%
TF molecules bind to their genomic binding sites by a combination of 3D diffusion through the cytosol and 1D random walk along the DNA (the \emph{facilitated diffusion} search mechanism). 
This mechanism was first proposed by \citet{riggs_1970a} to explain the fact that the \emph{lac} repressor (\emph{lacI}) in \emph{E. coli} finds its target site much more quickly than it would be possible by simple diffusion through the cytoplasm. It was later formalised by \citet{berg_1981} who found that it could indeed explain the reduced search time. 1D diffusion along the DNA, so called sliding  \footnote{Since it is difficult to experimentally distinguish between the 1D translocation modes, by sliding we refer to 1D random walk which includes both sliding and hopping}, was first shown \textit{in vitro} by \citet{kabata_1993}, but its significance \emph{in vivo} was disputed for a long time. Recently, using fluorescently tagged \emph{lac} repressor molecules, \citet{hammar_2012} directly observed TF sliding in living \textit{E. coli}. 

In order to calculate the average target search time of a TF using facilitated diffusion \citet{mirny_2009} use a model that includes alternating 3D diffusion and sliding events. They note that increasing the average number of different base pair positions visited during a sliding event, called the sliding length, has two adverse effects on the average search time: it decreases the number of slides needed to find the site, but it also increases the duration of a single slide, because more base pair positions have to be visited. It was then shown that the search time is minimal when the TF spends an equal amount of time sliding and using 3D diffusion during its search. 

Interestingly, \citet{elf_2007} found that \emph{lacI} spends about $90\%$ of its total search time sliding on the DNA, which differs significantly from the value that minimises target search time. It was suggested that, on crowded DNA, the observed fraction minimises the search time \citep{benichou_2011}. Another explanation for this would be that more time spent on the DNA optimises the system with respect to other properties. Indeed, previous work mostly assumes that the evolutionary advantage of facilitated diffusion is only due to the accelerated target search time, which could help to change gene expression more quickly in response to certain stimuli and signals. Other effects of TF sliding have rarely been investigated. 

In this analysis, we investigate another aspect of facilitated diffusion, namely how TF binding and unbinding in steady state affects gene expression noise of the controlled gene. In particular, we ask if TFs using facilitated diffusion lead to different gene expression noise when compared to an equivalent non-sliding TF and if this could provide a new view on the evolution of facilitated diffusion.  Furthermore, we also investigate how facilitated diffusion affects the activity changes of a controlled gene in steady state. Stochastic gene expression models often simply assume that genes switch between active (when the gene can be transcribed) and non-active states (when the gene cannot be transcribed) with constant stochastic rates. Here we try to evaluate how gene switching should be modelled for genes that are controlled by a TF using facilitated diffusion. 

Our results show that the facilitated diffusion mechanism can lead to a reduction in the fluctuations of mRNA and protein levels which is caused by its acceleration of target finding.  In addition, we found that, for biologically relevant parameters, the binding process can be represented by a two-state Markov model (if the effective binding and unbinding rates are chosen appropriately).

\section{Materials and Models}
\begin{figure}
\centering
\includegraphics[width=0.49\textwidth]{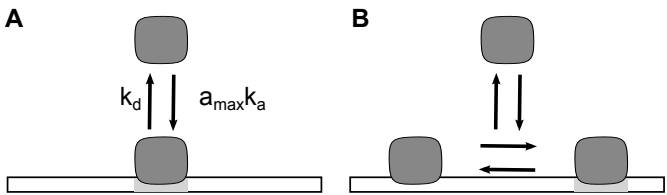}
\caption{\emph{Model of TF binding}. \onel\ shows binding and unbinding of TFs that are unable to slide (two-state Markov model).  \twol\ shows the binding dynamics of a TF that is able to slide on the DNA.}
\label{fig:modelCartoon}
\end{figure}

We consider two models, namely: \one the TF molecules perform only 3D diffusion and \two the TF molecules perform facilitated diffusion. In the former, when the molecule is bound to the target site, the TF has a constant rate of unbinding $k_d$, and the rate of rebinding $k_a$ for an individual TF can also be assumed to be constant \citep{redner_2001,zon_2006,mirny_2009}; see \fig\ \ref{fig:modelCartoon}\onel. In the case of multiple TF copies, the (re)binding rate is simply scaled up by the number of TFs per cell, $a_{\textrm{max}}$.

In the second model (the facilitated diffusion model), the TF molecules can slide off the target with a strongly increased chance of quickly sliding onto the target again; see \fig\ \ref{fig:modelCartoon}\twol. Hence the rebinding rate is not constant and binding cannot be modelled as a simple two-state Markov process as it is the case for non-sliding TFs. This binding mechanism can lead to long periods of no binding, when the TF diffuses through the cytoplasm, interrupted by short periods of multiple consecutive target binding events when the TF slides near the target site. 
In order to simulate the resulting expression of the controlled gene in steady state, we derived a stochastic model of TF binding and unbinding in case of facilitated diffusion.

\begin{figure}
\centering
\includegraphics[width=0.49\textwidth]{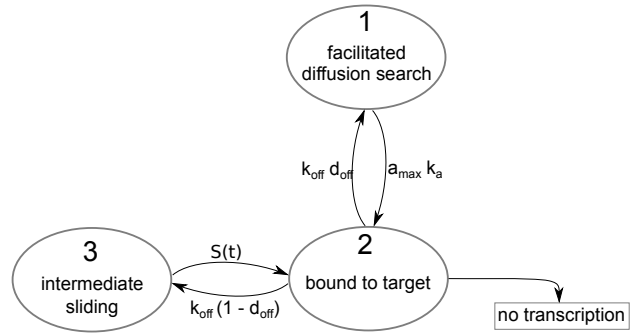}
\caption{\emph{A three-state system modelling the target binding dynamics of a TF using facilitated diffusion}. Each TF switches stochastically between: searching for the target combining 3D and 1D diffusion (state 1), being bound to the target (state 2) and sliding near the target between two consecutive target binding events (state 3). Note that $k_{\textrm{off}}$ represents the rate of leaving the target site, $d_{\textrm{off}}$ the probability of detaching from the DNA before returning to the target, $a_{\textrm{max}}$ the number of TF molecules per cell, $k_a$ the association rate of one free TF and  $S(t)$ the  probability density of sliding back to the target site after a time $t$. }
\label{fig:model3states}
\end{figure}

When the TF unbinds from the target site it can either start sliding along the DNA near the target and then slide back to it again or dissociate from the DNA strand before rebinding the target (probability $d_{\textrm{off}}$). These dynamics can be represented by a three-state model, where the TF is either: using combined 3D and 1D facilitated diffusion to search for the target (state 1), bound to the target (state 2) or sliding near the target between two consecutive target binding events (state 3); see \fig\ \ref{fig:model3states}. Each TF molecule stochastically switches between these states according to specific waiting time distributions. 
The transition rates are constant and the waiting times exponentially distributed, except in the case of switching from state 3 to state 2. This rate is not constant because first passage times in 1D diffusion are strongly distance dependent \citep{redner_2001}. Right after sliding off the target, the TF will still be close and have a high chance of rebinding, but after a long time without rebinding the probability of rebinding is much lower. Therefore this distribution of the waiting times decays faster than exponentially. 

To compute the shape of this waiting time distribution, we assume that the TF performs an unbiased continuous time random walk on the DNA with a step size of $1\ bp$ (see \citep{zabet_2012_review} for a discussion of these assumptions). Waiting times to slide over $1\ bp$ are exponentially distributed and all positions apart from the target site have the same mean waiting time $\Delta \tau$ (this holds for biologically relevant parameters \citep{ezer_2014}). Finally, when sliding near the target, there is a constant chance of unbinding from the DNA strand, with $\tau$ being the average time until unbinding.  

The probability density $S(t)$ of sliding back to the target site after a time $t$ is given by
\begin{equation}
S(t) =   D(t) \cdot F(t) 
\end{equation}
where $D(t)$ is the probability that the TF is indeed still bound to the DNA at time $t$ and $F(t)$ is the first return probability density of a continuous time random walk. $S(t)$ can then be calculated as
\begin{widetext}
\begin{equation}
S(t; \tau, \Delta \tau) =  e^{-t/\tau} \cdot \sum_{m = 1}^{\infty} \frac{2m \cdot e^{-\frac{t}{\Delta \tau}} \left (\frac{t}{ \Delta \tau} \right )^{2m-1}}{\Delta \tau \cdot (2m)!} \frac{2}{2m-1} \binom{2m-1}{m} 2^{-2m}
\end{equation}
\end{widetext}
; see \sm\ \ref{sec:appendixWaitingTimeDistribution}. Given that $d_{\textrm{off}}$ is the probability of detaching from the DNA before returning, we can write
\begin{equation}
d_{\textrm{off}} = 1 - \int\limits_0^\infty S(t) dt
\end{equation}

Defining the sliding length $s_l = \sqrt{2 \tau / \Delta \tau}$ as the average number of different base pair positions the TF visits during one slide \citep{wunderlich_2008}, we find that $d_{\textrm{off}} = 2/s_l$, which matches the results derived in \citep{hammar_2012}. The normalised distribution of $S(t)$ gives the waiting time distribution of switching from state 3 to state 2.

\section{Results}
\subsection{Evaluating the \emph{lac} repressor system}
To evaluate the two models with a biologically relevant set of parameters, we used experimental data from \emph{lacI} in \emph{E. coli}, which is a well characterised system; see \tab\ \ref{tab:3stateModelParameters}. We use the following notation: $b\in \{0,1\}$ is the number of TF molecules bound to the target site, $m$ the number of mRNA molecules in the cell and $p$ the protein level. In our model, we assumed that a repressor binding to the target would make transcription impossible and fully silence the gene.  When no TF is bound, single mRNA copies are produced at a constant rate ($\lambda_m$). The model also assumes that mRNA levels decay exponentially with rate $\beta_m$.

Investigating the three-state model using these parameters we found that, on average, the time spent sliding between two consecutive target binding events (state 3) is much shorter than both the time scale the TF is bound to the target and the time scale of transcription. This implies that the fast switching between target binding and intermediate sliding could be well represented by a single long binding event. It is important to note that the number of target binding events before dissociation from the DNA is not fixed but geometrically distributed. The total binding time before dissociating from the DNA strand is therefore given by the sum of a geometrically distributed number of exponential waiting times, which has the same distribution as a single long exponential waiting time; see \sm\ \ref{sec:appendixExponentialWaitingTime}. This means that TF binding patterns in case of fast enough sliding can be represented by a simple two-state system (search state and target binding state) with constant switching rates. 

In case of \emph{lacI}, we simulated both the three-state \emph{lacI} binding model and the resulting mRNA dynamics using a standard stochastic simulation algorithm \citep{gillespie_1977}. The algorithm was slightly adapted to correctly simulate the non constant target return rate from the intermediate sliding state; see \sm\ \ref{sec:appendixSSA}. Similarly we simulated the dynamics of the corresponding two-state system where multiple returns due to sliding are combined to a single continuous binding event. To obtain the same binding time, we set the rate of unbinding from the target in the two-state system equal to the unbinding rate in the three-state system divided by the average number of target returns before detaching from the DNA.
\begin{equation}
k_d = k_{\textrm{off}}\cdot d_{\textrm{off}}=2\cdot k_{\textrm{off}}/s_l \label{eq:rateChange}
\end{equation}

\begin{figure}
\centering
\includegraphics[angle=270,width=0.49\textwidth]{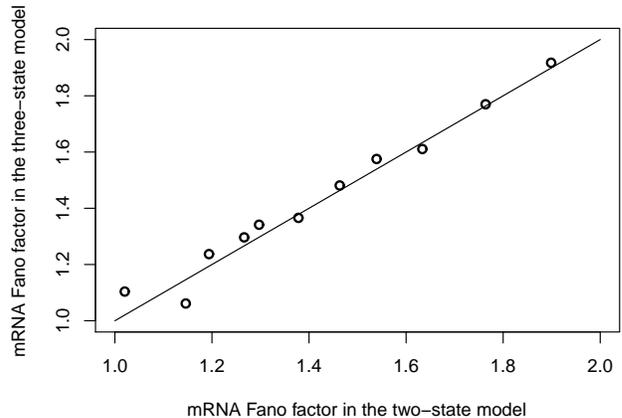}
\caption{\emph{Comparison of the mRNA noise in the two models}. The diagonal indicates that the two models produce similar results. Additional data points show the same comparison in the case of TF that are similar to \emph{lacI} but have slightly different sliding and/or binding rates (up to a 10 fold difference). The circles represent the average Fano factor computed over 10 stochastic simulations.}
\label{fig:modelsComparisonPlot}
\end{figure}

We found that in both scenarios the average mRNA level is $0.16$ molecules per cell. \fig\ \ref{fig:modelsComparisonPlot} shows the mRNA Fano factor in the two-state and three-state system for \emph{lacI} as well as hypothetical TFs with up to 10 times slower sliding and up to 10 times faster or slower binding/unbinding rates. In each case, the difference between the Fano factor computed using the two models is negligible. This indicates that, for TFs with similar dynamics as \emph{lacI}, the gene regulation process can be appropriately modelled by the two-state model (with constant binding and unbinding rate), which is supported by previous work that successfully modelled experimentally measured lac mRNA noise using a two-state Markov model \citep{so_2011}.

 Further discussion of model assumptions and comparison to relevant previous work can be found in \sm\  \ref{sec:AppendixModelConsiderations}.

\subsection{Effect of faster target finding rate on mRNA fluctuations}
Next, we investigate if the speed up in target finding due to facilitated diffusion significantly affects the steady state fluctuations in the \emph{lacI} mRNA levels compared to a non-sliding equivalent. Here TF sliding was not taken into account explicitly any more, rather we used the equivalent two-state Markov model with effective binding rates derived previously to model TF binding. In order to allow a sensible comparison between these two TFs, we required that, in both cases, the TF is bound to the target the same fraction of the time such that the average level of mRNA is the same. If there are $a_{\textrm{max}}$ TF in a cell, each of which binds the target at a constant rate $k_a$ and unbinds at a rate $k_d$, the average fractional time $b$ the TF is bound to the target is  \citep{chu_2009}
\begin{equation}
b = \frac{k_a a_{\textrm{max}}}{k_a a_{\textrm{max}} + k_d}
\end{equation}
Assuming that in both cases the TF number per cell and hence the metabolic cost is the same, having identical binding times $b$ also requires the ratio $k_a/k_d$ to be the same in both cases, i.e., a slower target finding rate has to be compensated by an equal decrease in the unbinding rate (see discussion in \sm\  \ref{sec:AppendixModelSildingNonSliding} for specific details about the comparison between sliding and non-sliding TFs).  

To compare mRNA fluctuations for \emph{lacI} to its non-sliding equivalent, we considered the two-state Markov model, where the mRNA Fano factor can be derived analytically \citep{paulsson_2005} as
\begin{equation} \label{eq:twoStateProteinVariance}
\frac{\sigma^2_m}{m} = 1 + m \frac{b}{1-b} \frac{\tau_b}{\tau_b + \tau_m} 
\end{equation}
with $b$ being the average fractional time the TF is bound, $m$ the average mRNA level, $\tau_b = (k_a + k_d)^{-1}$ the time scale of gene switching and $\tau_m$ the time scales of mRNA degradation. There are two sources of noise in the mRNA level  \citep{paulsson_2005}: \one the intrinsic Poisson noise arising from the stochastic nature of each transcription and mRNA degradation event and \two the extrinsic component arising from the random switching of the gene's activity. 

\begin{figure}
\centering
\includegraphics[angle=270,width=0.49\textwidth]{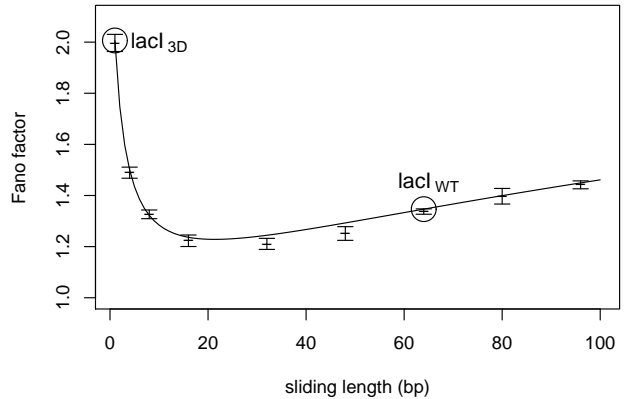}
\caption{\emph{Dependence of the mRNA Fano factor on sliding length.} The dissociation rate was changed accordingly to keep average mRNA levels constant. Analytic results (line) were calculated using equation \eqref{eq:twoStateProteinVariance}. For each set of parameters, we also performed $10$ stochastic simulations each run over $2000$ reaction events (the error bars are $\pm s.d.$). Fano factors for lacI  (\emph{lacI}$_{\textrm{WT}}$) and an equivalent TF that does not slide (\emph{lacI}$_{\textrm{3D}}$) are highlighted specifically. }
\label{fig:mRNAnoise}
\end{figure}

Using equation \eqref{eq:twoStateProteinVariance} and the appropriate parameters (see \tab\ \ref{tab:3stateModelParameters}) the \emph{lac} operon mRNA Fano factor in steady state was found to be $\sigma^2_{m,lacI}/m = 1.3$. By setting the sliding length to $s_l = 1$, the binding rate of the 3D diffusion \emph{lacI} equivalent  is $k_a^{3D}=k_a/6.4$; see \sm\ \ref{sec:appendixAssocRate}. Decreasing both $k_a$ and $k_d$ by this factor, we found that the mRNA Fano factor is $\sigma^2_{m,3D}/m = 2.0$. The accelerated target finding due to facilitated diffusion therefore leads to a noise reduction of $33\%$ in case of \emph{lacI}. \fig\ \ref{fig:mRNAnoise} shows the levels of mRNA fluctuations for \emph{lacI}, when assuming various sliding lengths. Since we adjusted the dissociation rate to ensure equal average expression levels (see \sm\ \ref{sec:AppendixModelSildingNonSliding}), here the mRNA noise is solely determined by the target finding rate, i.e. faster target finding directly leads to a lower Fano factor. Note that sliding lengths that are slightly shorter than the wild type value show the lowest mRNA Fano factor. This is due to the fact that \emph{lacI} is bound to the DNA about 90\% of the time \citep{elf_2007}, while fastest binding would be obtained at a value of 50\% and hence at lower sliding lengths \cite{slutsky_2004b,mirny_2009}. For sliding lengths that are longer than the wild type value, the fraction the TF is bound to the DNA is even higher, leading to slower target finding rates and, consequently, higher mRNA noise levels. 

\subsection{Effects on protein noise}
To quantify the fluctuations in the protein level ($p$), we added two reactions to the previously used reaction system - each mRNA molecule is translated at a constant rate ($\lambda_p$), while the resulting proteins are degraded exponentially (decay rate $\beta_p$). Parameter values were taken from $\beta-galactosidase$ measurements; see \tab\ \ref{tab:3stateModelParameters}.

\begin{figure*}
\centering
\includegraphics[angle=270,width=\textwidth]{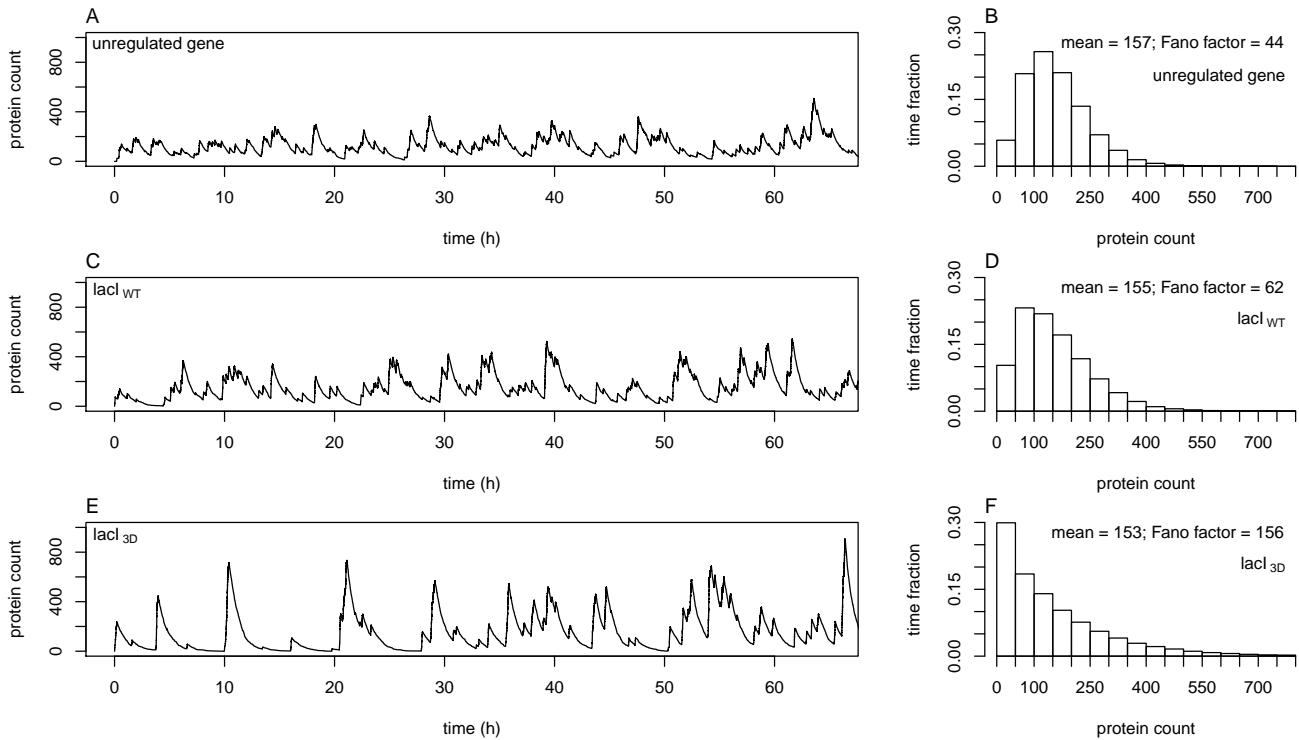}
\caption{\emph{Protein fluctuations}. We computed the protein counts of $\beta-galactosidase$ for three different cases: ($A-B$) the gene is constantly \emph{on} (unregulated gene), ($C-D$) the gene is regulated by \emph{lacI}  ($lacI_{WT}$) and ($E-F$) the gene is regulated by \emph{lacI}-like TF that does not slide on the DNA ($lacI_{\textrm{3D}}$). Each system was simulated over a real time equivalent of $72\ h$. ($B,D,F$) Each histogram uses the data from a simulation using $10^6$ reactions. }
\label{fig:proteinNoise}
\end{figure*}

\fig\ \ref{fig:proteinNoise} shows the simulated fluctuations in protein level in three different cases: \one gene is permanently \emph{on}, \two gene is controlled by the \emph{lacI} and \three gene is controlled by the non-sliding \textit{lacI} equivalent. In each system, the transcription rate is set to a value so that the average protein level is $\langle p \rangle = 150\ molecules$. The case in which the gene is permanently \emph{on} shows the weakest fluctuations, whereas protein levels fluctuate most strongly in the 3D diffusion case. Facilitated diffusion reduces the fluctuations in protein levels compared to the case of the TF performing only 3D diffusion, but it cannot reduce it under the levels of an unregulated gene.

\section{Discussion}
%
Gene expression is a noisy process \citep{even_2006,newman_2006,taniguchi_2010} and,  in order to understand the gene regulatory program of cells, it is important to investigate its noise properties. Usually it is assumed that genes get switched on and off due to binding and unbinding of TFs and when they are \emph{on} they are transcribed at constant rates. 
The resulting mRNAs then are translated also at constant rates. 
One aspect that is often neglected in this model is that the binding of TFs to their binding sites is not a simple two-state Markov process, but rather TFs perform facilitated diffusion when binding to their binding sites. In this contribution, we investigated how this model of binding of TFs to their target sites affects the noise in gene expression. 

First, we constructed a three-state model that is able to describe the dynamics of TFs when performing facilitated diffusion; see \fig\ \ref{fig:model3states}. Our results show that, in the case of TFs that slide fast on the DNA, the noise and steady state properties of the three-state model of TF binding to their target site can be described by a two-state Markov model, when the unbinding rate of the two-state model is set to $k_d = 2\cdot k_{\textrm{off}}/s_l$ (this is similar to the result in \citep{zon_2006}, which considered only hopping and no sliding); see \fig\ \ref{fig:modelsComparisonPlot}. Interestingly, DNA binding proteins seem to move fast on the DNA when they perform a 1D random walk (e.g. see \tab\ I in \citep{desantis_2011}) and this suggests that, when modelling TF binding to their binding site, the assumption of a simple two-state Markov process does not introduce any biases. We specifically show that this is the case when parameterising our model with experimental data from the lac repressor system. It is worthwhile noting that, both in bacteria and eukaryotes, the two-state Markov model seems to accurately account for the noise in gene regulation \cite{so_2011,taniguchi_2010,sanchez_2013}, but there are also exceptions where the kinetic mechanism of transcription is encoded by the DNA sequence, for example gene expression in yeast \cite{sanchez_2013} or \emph{eve stripe 2} expression in \emph{D. melanogaster} \citep{bothma_2014}.    

This indicates that the effect of facilitated diffusion on gene expression noise is limited to changing the effective constant binding and unbinding rates of the TF. We investigated how the increased target finding rate due to facilitated diffusion changes the noise in case of \emph{lacI}. Our results show that non-sliding TFs (with the same 3D diffusion coefficient, average target binding times and identical per cell abundance as \textit{lacI}) lead to a strongly increased noise in both mRNA (see \fig\ \ref{fig:mRNAnoise}) and protein levels (see \fig\ \ref{fig:proteinNoise}) when compared to equivalent TFs that slide. This suggests that, in addition to the increase in speed of binding of TFs, facilitated diffusion could also lead to lower noise. Experimental studies found that, in \emph{E. coli}, the mRNA noise is correlated with the mRNA levels and that TF binding kinetics do not seem to have a strong contribution to mRNA noise \citep{taniguchi_2010,so_2011}. Our results suggest that one potential explanation for this result is that facilitated diffusion buffers this noise in gene regulation. In other words, when assuming that TFs perform facilitated diffusion, the contribution of the binding/unbinding kinetics to the mRNA/protein noise is relatively small; see \fig\ \ref{fig:mRNAnoise} and \fig\ \ref{fig:proteinNoise}. 

It is important to note that increasing the number of non-sliding TFs per cell by a factor of 6.4 leads to the same acceleration in target finding and hence to equally low expression fluctuations, but also a higher metabolic cost. The facilitated diffusion mechanism is able to reduce the noise and response time of a gene without increasing the metabolic cost of the system \citep{zabet_2009} and without increasing the complexity of the promoter (by adding auto-repression) \citep{zabet_2011}.

Our model assumes a naked DNA although \emph{in vivo} it would be covered by other molecules.  In \citep{zabet_2013_time}, we performed stochastic simulations of the facilitated diffusion mechanism and found that molecular crowding on the DNA can increase the noise in gene regulation, but at biological relevant crowding levels, this increase is small. This result can be explained by the fact that molecular crowding on the DNA reduces the search time, but this reduction is not statistically significant \citep{zabet_2013_time,brackley_2013_crowding}.     

Furthermore, our model also assumes that there are no other nearby binding sites, which could potentially affect the results \citep{hammar_2012,ezer_2014,sharon_2014}. Recently, Sharon \emph{et al.} \cite{sharon_2014} showed that synthetic promoters consisting of homotypic clusters of TF binding sites can lead to higher noise and this noise is accounted by the fact that TFs perform facilitated diffusion. However, in case of \emph{lacI}, the binding site that is closest to $O_1$ is further away than its sliding length, thus confirming the validity of our findings. For the case of densely packed promoters, the influence of facilitated diffusion on noise in gene expression needs a systematic investigation, but this will  be left to future research.

We would also like to mention that although all relevant parameters were taken from the lac repressor system, several aspects of the system (such as the cAMP-bound catabolite activator protein) have been neglected. Due to these limitations the model cannot be used to fully describe the lac operon behaviour. Instead parameters from the lac system are used to evaluate our model within a biologically plausible regime. Despite the abstraction level of our model, for the $P_{lac}$ system, we predicted a mean mRNA level of about 0.16 per cell and assuming facilitated diffusion, we estimated the Fano factor to be $1.3$ (as opposed to $2.0$ in the case of TF performing only 3D diffusion), which is similar to the values measured experimentally in the low inducer case by \citep{so_2011} (for $\langle m\rangle \approx 0.15$ the Fano factor is $\approx 1.25$). Our results suggest that facilitated diffusion is essential in explaining the experimentally measured noise in mRNA and that one does not need to model the 1D random walk explicitly, but rather include the effects of facilitated diffusion in the binding rate. Further validation of our model would consist of  changing the sliding length of a TF by altering its non-specific interactions (see for example \cite{vuzman_2010_tails_composition,tafvizi_2011}) and then measuring the gene expression noise in these systems. However, it is not clear how these changes will affect the capacity of the TF to regulate the target genes and a systematic analysis is required to investigate these additional effects.

\appendix
    \begin{center}
      {\bf APPENDIX}
    \end{center}

\section{Waiting time distribution when sliding back to the target before unbinding the DNA} \label{sec:appendixWaitingTimeDistribution}

The chance that a TF slides back to the target at a time $t$ after it slid off it, $S(t)$, is given by the probability of first return to the origin after time $t$ during a simple unbiased continuous time random walk $F(t)$, times the probability that the TF is still bound to the DNA at time $t$, $D(t)$.

\begin{equation}
S(t) = D(t) \cdot F(t) 
\end{equation}

 The probability that the TF is still bound to the DNA at a time $t$ after unbinding the target decays exponentially with characteristic waiting time $\tau$, i.e. $D(t;\tau) = e^{-t/\tau}$.

Since $F(t)$ is the probability density function of first return to the origin at time $t$, it is given by the probability of first return after $n$ $1\ bp$ steps, $F_n$, multiplied by the probability density of making the $n^{th}$ step at time $t$, $\phi_n (t)$, and then marginalising over all $n$:

\begin{equation}
F(t) = \sum_{n = 1} ^{\infty} F_n \cdot \phi_n(t)
\end{equation}

According to \citet{klafter_2011}, these probabilities can be calculated to be 

\begin{equation}
F_n = \frac{2}{n-1} \binom{n-1}{n/2} 2^{-n} \text{, for even $n$ and $0$ otherwise}
\end{equation}

and 

\begin{equation}
\phi_n(t) = \mathcal{L}^{-1} \left \{\phi^n(s) \right \}
\end{equation}

the inverse Laplace transform of $\phi^n(s)$, where $\phi(s)$ is in turn the Laplace transform of the waiting time distribution for a single base pair step, $\phi(t)$. Here we assume that the waiting time of sliding one step in the neighbourhood of the target is exponentially distributed with a constant characteristic time scale $\Delta \tau$. Therefore, $\phi(s) = \frac{1}{1+ s \Delta \tau}$ and $\phi_n(t)$ can be calculated to be

\begin{equation}
\phi_n(t;\Delta \tau) = \frac{n e^{-\frac{t}{\Delta \tau}} \left (\frac{t}{ \Delta \tau} \right )^{n-1}}{\Delta \tau \cdot n!}
\end{equation}

Since $F_n$ vanishes for odd $n$, we can get $F(t)$ by summing over all $n = 2m$ to yielding the following expression for the rebinding time distribution: 

\begin{widetext}
\begin{equation}
S(t; \tau, \Delta \tau) = e^{-t/\tau} \cdot \sum_{m = 1}^{\infty}  \frac{2m \cdot e^{-\frac{t}{\Delta \tau}} \cdot \left (\frac{t}{ \Delta \tau} \right )^{2m-1}}{\Delta \tau \cdot (2m)!} \frac{2}{2m-1} \binom{2m-1}{m} 2^{-2m}
\end{equation}
\end{widetext}

Note that this distribution is not normalised, since the probability of sliding back to the target before unbinding from the DNA is smaller than 1. However, the waiting time in state 3 in the TF binding model is the probability density of returning at time $t$ given that it does return before unbinding the DNA. The waiting time therefore has to be drawn from the corresponding normalised distribution of $S(t; \tau, \Delta \tau)$.

Our model assumes that unbinding directly from the target site is negligible. If a TF molecule performs $s_l^{2}/2$ events during a 1D random walk and the probability to unbind is equal from all positions, then the probability to unbind during any of these events is $2/s_l^{2}$ \citep{zabet_2012_model}. Given that on average a TF molecules visits the target site $s_l/2$ times during a 1D random walk, then the probability to dissociate directly from the target site is $1/s_l$, which for our model is less than $1.5\%$ and, thus, was neglected here.

\section{Geometrically distributed number of returns leads to an overall exponentially distributed target binding time}\label{sec:appendixExponentialWaitingTime}

In case of sufficiently fast sliding, TFs moving on and off the target multiple times can be approximated by a single long target binding event. The length of this effective binding event is given by the sum of all individual binding events. Here each individual binding time is exponentially distributed. The number of consecutive binding events before DNA detachment is geometrically distributed since each time the TF leaves the target site there is a constant chance $d_{\textrm{off}}$ of not returning to the target through sliding. Here we derived the time distribution of the overall waiting time as a sum of a geometrically distributed number of exponential waiting times. 

The waiting time distribution of an individual binding event is 
\begin{equation}
\phi(t) = \frac{1}{\Delta \tau} e^{-t/\Delta \tau}
\end{equation}

The overall effective waiting time density function given that the TF binds the target exactly $n$ consecutive times is
\begin{widetext}
\begin{equation}
P(t|n) = \int\limits_0^{t_2}\int\limits_{t_1}^{t_3} \dots \int\limits_{t_{n-2}}^t \phi(t_1) \cdot \phi(t_2 - t_1) \cdot \dotsc \cdot \phi(t - t_{n-1})dt_1 dt_2 \dots dt_{n-1}
\end{equation}
\end{widetext}
In Laplace domain, these convolutions turn into a simple product.
\begin{equation}
P(s|n) = \left [ \phi(s) \right ]^n
\end{equation}
with $\phi(s) = \frac{1} {1+\Delta \tau s}$ being the Laplace transform of $\phi(t)$.

We assume that the number of individual binding events $n$ is geometrically distributed with constant chance $d_{\textrm{off}}$ of not sliding back. The joint probability is therefore
\begin{equation}
P(s,n) = P(s|n) \cdot (1-d_{\textrm{off}})^{n-1} \cdot d_{\textrm{off}}
\end{equation}

and hence the return time distribution is

\begin{eqnarray}
P(s) &=& \sum_{n = 1}^{\infty} \left [ \phi(s) \right ]^n \cdot (1-d_{\textrm{off}})^{n-1} \cdot d_{\textrm{off}}\nonumber\\
 &=& \frac{\phi(s)\cdot d_{\textrm{off}}} {1 - \phi(s) (1-d_{\textrm{off}})}
\end{eqnarray}

Substituting $\phi(s)$ from above

\begin{equation}
P(s) = \frac{d_{\textrm{off}}}{d_{\textrm{off}} + \Delta \tau s} = \frac{1}{1 + N \Delta \tau s}
\end{equation}

and 

\begin{equation}
P(t) = \frac{1}{N \Delta \tau} e^{-t/N \Delta \tau}
\end{equation}

where $N = 1/d_{\textrm{off}}$ is the average number of target bindings before DNA unbinding. We can conclude that in case of fast enough sliding, multiple returns to the target can be modelled as a single binding event that is exponentially distributed with average binding time $N\cdot\Delta\tau$.

\section{Change to the stochastic simulation algorithm} \label{sec:appendixSSA}

The stochastic simulation algorithm used by \citet{gillespie_1977} appropriately simulates reaction systems with exponential waiting times, i.e. systems with all possible reactions occurring at constant rates for a specific configuration. This is the case for all reactions in our system apart from the TF sliding back to the target site. When the TF slides off the target the return rate is not constant but decays with time. 

In order to appropriately simulate our system, we slightly adapted the stochastic simulation algorithm. The original algorithm draws the time of the next reaction from an exponential distribution with a rate equal to the sum of all possible reactions in the current configuration. Then the specific reaction is chosen according to the individual rates. Here we do the same for all constant rate reactions in the system, but, in case of the TF being in the sliding state, we additionally draw a waiting time from the return time distribution $S(t)$, derived earlier. If the waiting time drawn from $S(t)$ is smaller than the other, the TF returns to the target. If not, a constant rate reaction is carried out accordingly.

\section{The parameters of the three state model}
The list of parameters for the three state model are listed in Table \ref{tab:3stateModelParameters}. Below, we described how some of the parameters were derived.

\begin{table}
\centering
\begin{tabular}{| c | r | r |}
\hline
parameter & value & reference\\ \hline
$a_{\textrm{max}}$ & $5\ molecules$ & \citep{gilbert_1966}\\ \hline
$s_l$ & $64 \pm 14\ bp$  & \citep{hammar_2012} \\ \hline
$k_{\textrm{a,FD}}$ & $(0.0044\pm0.0011)\ s^{-1}$ & \citep{hammar_2012,li_2009} \\ \hline
$k_{\textrm{d}}$ & $ 0.0023\ s^{-1} $ & \citep{fried_1981} \\ \hline
$k_{\textrm{off}}$ & $ 0.074\ s^{-1} $ & equation \eqref{eq:koff} \\ \hline
$\tau$ & $5\ ms$ & \citep{elf_2007} \\ \hline
$\Delta\tau$ & $2.4\ \mu s$ & \citep{elf_2007,hammar_2012}\\ \hline
$\lambda_m$ & $0.012\ s^{-1}$ & \citep{malan_1984,golding_2005} \\ \hline
$\beta_m$ & $(0.007\pm0.001)\ s^{-1}$  & \citep{kennell_1977,selinger_2003,ehretsmann_1992} \\ \hline
$\lambda_p$ & $0.32\ s^{-1}$ & \citep{mandelstam_1957} \\ \hline
$\beta_p$ & $0.0033\ s^{-1}$ & \citep{kennell_1977} \\ \hline
\end{tabular} 
\raggedright
\caption{\emph{Parameter values}. }
\label{tab:3stateModelParameters}
\end{table}

\subsection{Number of \emph{lacI} operons per growing \emph{E. coli} cell}

Although the lac operon only occurs once in the \emph{E. coli} genome \citep{riley_2006}, continuous DNA replication during growth can lead to more than one gene being present in a growing cell. Usually, one could observe only one binding spot for \emph{lacI}, when investigating $lacI$ binding in living and growing cells \citep{elf_2007}. Thus, we assumed that there is only about one lac operon present in each growing \emph{E. coli} cell.

\subsection{Total number of \emph{lacI} molecules per cell}

There are $20$ \emph{lacI} monomers per \emph{lacI} gene in wild type \emph{E. coli} \citep{gilbert_1966} and, since there is only one gene per cell (see above), we estimate that there are only  $a_{\textrm{max}} = 5$ independently searching lac tetramers per cell. 

\subsection{Sliding length $s_l$}

The root mean square deviation during one slide on the DNA was estimated to be $s_{l,RMSD} = \sqrt{2 D_{\textrm{1D}} / k_d} = (45 \pm 10)\ bp$ \citep{hammar_2012}, where $D_{\textrm{1D}}$ is the 1D diffusion constant and $k_d$ is the DNA dissociation rate. Here, we defined the sliding length $s_l$ as the average number of different base pairs that the TF visits at least once during one slide. Thus, we can compute the sliding rate as $s_l = \sqrt{4 D_{\textrm{1D}} /k_d} = \sqrt{2} s_{l,RMSD}$ \citep{elf_2007,wunderlich_2008} and, thus, $s_l = (64 \pm 14)\ bp$. \citet{hammar_2012} does not discuss if this sliding length includes short dissociation events followed by immediate rebinding (hopping) or if the TF unbinds the first time on average after scanning $64\ bp$ with a chance of immediately binding again, performing a new slide on the DNA. The experimental approach used to determine the sliding length \citep{hammar_2012} consisted of  measuring how the association rate decreases as additional binding sites near the target are introduced. Given a median hopping distance of $1\ bp$ and about $6$ hops per 1D random walk \citep{wunderlich_2008}, it is very unlikely that hops would by chance overcome the extra binding site. Hopping is therefore unlikely to significantly alter the experimental results suggesting that $s_l = (64 \pm 14)\ bp$ already includes short hops.

\subsection{The dissociation rate from the binding site}

The dissociation rate from the binding site is computed using the following equation from the main text
\begin{eqnarray}
 & & k_{\textrm{d}}=\frac{2\cdot k_{\textrm{off}}}{s_l}\quad \Rightarrow \quad \nonumber\\
 & & k_{\textrm{off}} =\frac{s_l\cdot k_{\textrm{d}}}{2} =\frac{64\cdot  0.0023}{2}= 0.074\ s^{-1} \label{eq:koff}
\end{eqnarray}
Note that we used the following values: $s_l=64\ bp$ (see above) and $k_{\textrm{d}}=0.0023\ s^{-1}$ \citep{fried_1981}.  The latter is similar to the value measured recently (mean bound time of $5.3 \pm 0.2$) using a single molecule chase assay \citep{hammar_2014}.

\subsection{$\beta-galactosidase$ translation rate}

\citet{kennell_1977} measure one translation initiation of a single lacZ mRNA every $2.2\ s$ in exponentially growing cells. However they state that around $30\%$ of the polypeptides are not completed, giving one effective translation every $3.1\ s$ and a effective translation rate of $\lambda_p = 0.32\ s^{-1}$. 

\subsection{$\beta-galactosidase$ protein decay rate}

\citet{mandelstam_1957} measured a $\beta-galactosidase$ degradation rate of $1.4\cdot 10^{-5}\ s^{-1}$. This is much slower than the average protein dilution rate of an exponentially growing \textit{E. coli} cell of $3.3\cdot 10^{-4}\ s^{-1}$ \citep{bremer_1996}. Thus, the decay of $\beta-galactosidase$ is dominated by dilution and we approximate it by $\beta_p = 3.3\cdot 10^{-4}\ s^{-1}$.

\section{Changing to association rate to a non-sliding equivalent TF} \label{sec:appendixAssocRate}

Variations in the extent of facilitated diffusion during target finding can be achieved by varying the sliding length. This hypothetical TF, similar to $lacI$ in all respects but the sliding length, will have modified association rates. The association rate can be calculated in closed form, as outlined below.
The association rate $k_{a,s_l}$ of a TF with sliding $s_l$ is given by the following expression \citep{mirny_2009}
\begin{equation} \label{eq:associationRateLocal}
k_{a,s_l} = \frac{s_l}{M^*} (t_{1D,sl} + t_{3D})^{-1}
\end{equation}
where $M^*$ is the number of accessible base pairs in the genome and $t_{1D,sl}$ and $t_{3D}$ are the average durations of 1D searches (slides and hops on the DNA) and 3D searches (free diffusion in the cytoplasm). It has been experimentally observed that $lacI$ spends about $90\%$ of the time sliding when searching for the target site \citep{elf_2007}, which means that 
\begin{equation} \label{eq:t1Dt3D}
t_{1D,lacI} = 9\cdot t_{3D}
\end{equation}

To find the dependence of the association rate of the TF on the sliding length from equation \eqref{eq:associationRateLocal}, we need to calculate the modified $t_{1D,sl}$ and $t_{3D}$. Since the 3D search round duration is not affected by the sliding length of the TF, $t_{3D}$ is identical to that of $lacI$ and can be calculated by inverting equation \eqref{eq:associationRateLocal}:
\begin{equation} \label{eq:t3D}
t_{3D} = \frac{s_{l,lacI}}{10 M^* k_{a,lacI}}
\end{equation}

The average time spent during the 1D slide, $t_{1D,sl}$, is proportional to the average number of 1bp sliding steps $N$ performed during such a slide. Also, since the transcription factor diffuses along the DNA while sliding, $N$ is proportional to the square of $s_l$ \citep{wunderlich_2008} and $t_{1D,sl} \propto s_l^2$. Hence
\begin{equation} \label{eq:propto}
t_{1D,s_l} = t_{1D, lacI}\left( \frac{s_l}{s_{l,lacI}}\right)^{2}
\end{equation}

Combining equations \eqref{eq:associationRateLocal}, \eqref{eq:t1Dt3D}, \eqref{eq:t3D} and \eqref{eq:propto}, we find that the association rate of a TF with sliding length $s_l$:
\begin{equation} \label{eq:variableAssociationRate}
k_{a,s_l} = 10 k_{a,lacI}\frac{ s_l}{s_{l,lacI}} \left[ 9 \left( \frac{s_l}{s_{l,lacI}}\right)^{2} + 1 \right]^{-1}
\end{equation}
where $s_{l,lac} = (64 \pm 14)\ bp$ is the sliding length of \emph{lacI} \citep{hammar_2012} and $k_{a,lacI} = (0.0044 \pm 0.0011)\ s^{-1}$ is its association rate \citep{hammar_2012}. 

The association rate of an equivalent TF with a different sliding length can be found by plugging the sliding length $s_l$ into equation \eqref{eq:variableAssociationRate}. The 3D diffusion case can be approached by setting $s_l = 1\ bp$. In the 3D case, the reduced association rate is:
\begin{eqnarray}
k_{a,3D} &=& k_{a,lacI} \frac{10}{s_{l,lacI}} \left ( \frac{9}{s_{l,lacI}^2} + 1 \right )^{-1} \nonumber\\
 &=& k_{a,lacI} / 6.4 = 6.9 \cdot 10^{-4}\ s^{-1}
\end{eqnarray}
Hence, if $lacI$ was not using facilitated diffusion, it would take on average $6.4$ times longer find its target site.

\section{Further considerations on our model} \label{sec:AppendixModelConsiderations}
\subsection{Transcription initiation}
In our model, we do not model transcription explicitly, but we rather assume that an mRNA molecule is produced at exponentially distributed time intervals when the TF is not bound to the target site. Recently, \citep{hammar_2014} found that while this equilibrium model of transcription is accurate for certain promoters (including $lacO_1$), it fails to explain the behaviour of other promoters (e.g. $lacO_{sym}$). Nevertheless, these non-equilibrium binding mechanisms need systematic investigation and will be left to further research.

\subsection{Considerations on our three-state model}
In this contribution, we proposed a three-state model that described the facilitated diffusion mechanism. Pulkkinen and Metzler \citep{pulkkinen_2013} modelled facilitated diffusion analytically assuming a different three state model, namely, they assumed that the TF molecule can be in the following three states: \one free in the cytoplasm/nucleoplasm, \two bound non-specifically to the DNA in the vicinity of the target site and \three bound to the target site. The transitions between these three states were assumed to be exponentially distributed. 

Crucially, we considered that the TF molecule can be in different three states, namely: $1$ searching for the target using facilitated diffusion (at least one DNA detachment before target rebinding), $2$ bound to the target site and $3$ sliding on the DNA between two consecutive target binding events without DNA detachment. Note that when sliding off the target site, the TF molecule can be in both states $1$ and $3$, i.e., if it will return before DNA detachment, the TF is in state $3$, while otherwise in state $1$. Hence, we used well defined abstract states instead of a purely spatial definition as used in \citep{pulkkinen_2013}. In other words, we avoided a necessarily approximate definition of a ``local'' search state, which allows us to find the exact target return time distribution assuming facilitated diffusion of a TF. Importantly we find that when sliding on the DNA near the target site, the binding time is not exponentially distributed as it is assumed by \citet{pulkkinen_2013}. 

Furthermore, Meyer \emph{et al.} \citep{meyer_2012} investigated the noise in mRNA assuming that the search takes place in a compact environment and compared this with the case of the search taking place in a non-compact environment. They derived a non-exponential return rate to the target site and assumed that facilitated diffusion can be seen as a search in a compact environment. Our approach was different in the sense that we did not assume a distribution of the return times, but rather derived this distribution analytically by assuming a known model of facilitated diffusion. We further used this distribution and parameters derived from previous experiments to understand the influence of facilitated diffusion on the noise in mRNA and protein.

The main focus of our paper is what are the effects of facilitated diffusion on mRNA and protein noise. Pulkkinen and Metzler \citep{pulkkinen_2013} investigate this problem, but in the case of co-localisation of the gene encoding for a TF and the target site of that TF. This assumption makes their results valid only in the context of bacterial systems (where transcription and translation are co-localised), while our results are potentially valid even in the context of eukaryotic systems (where translation takes place outside the nucleus). Interestingly, it seems that mRNA noise in animal cells seem to display a similar level of correlation with the mean expression level as in the case of bacterial cells \citep{sanchez_2013}. This means that assuming that TFs perform facilitated diffusion in higher eukaryotes \citep{hager_2009,vukojevic_2010,tafvizi_2011,chen_2014}, the contribution from binding/unbinding kinetics is potentially small.

It is worthwhile noting that Pedraza and Paulsson \citep{pedraza_2008} proposed a general model to compute noise in mRNA where any distribution for the arrival times of the TFs to the target site can be assumed. Our model particularises this type of model to the case of facilitated diffusion and we explicitly derive the arrival time distribution as being non-exponential. 

Finally, we would like to emphasise that, to our knowledge, no previous work systematically compared non-sliding with sliding TFs and discussed the effects of facilitated diffusion on the noise in gene expression compared to simple 3D diffusion of TFs.

\subsection{Comparing sliding TFs to their hypothetical non-sliding equivalents}  \label{sec:AppendixModelSildingNonSliding}

van Zon \emph{et al.} \cite{zon_2006} investigated a different TF search effect, namely how fast rebinding in case of a TF that uses only 3D diffusion affects transcriptional noise. In our manuscript, we investigate the case of multiple returns due to sliding and find that facilitated diffusion leads to a reduction in the mRNA noise. This is different from the result of van Zon \emph{et al.} \cite{zon_2006}, who find that fast 3D diffusion returns increase transcriptional noise. The system investigated in our manuscript is different in that, unlike 3D diffusion returns, sliding does not only lead to multiple consecutive binding events but also leads to a speedup in target search, and hence increasing the TF target finding rate. Most importantly, the crucial difference between the two works that explains the seemingly contradictory conclusions is due to the difference in the questions posed. On one hand, van Zon \emph{et al.} \cite{zon_2006}  asked what happens to transcriptional noise if TFs quickly return to the target multiple times through 3D diffusion and hence decrease the effective dissociation rate. On the other hand, we ask how the effect on gene expression noise could pose an evolutionary advantage that could play a role in the development of facilitated diffusion. More specifically, we do not simply ask how a sliding TF compares to another TF that is identical, except that it is unable to slide along the DNA, but rather we investigate how the noise in gene expression in a system that has evolved using a sliding TF differs from the noise in gene expression in a system that uses a non-sliding TF. Thus, we require that both systems have the same average level of repression and this means that the average time a TF is bound to the target should be identical. 

Since we show that target binding dynamics of sliding TFs can be represented as an effective two state model, any possible advantage of the facilitated diffusion mechanism in terms of noise in gene expression must lie in the effective binding and unbinding rates. Here, we compared sliding and non-sliding TFs at equal TF number and thus, at equal metabolic cost. In the case of non-sliding TFs, the overall target finding rate is slower. To keep the average repression level the same, the target dissociation rate for the non-sliding TF is then decreased accordingly to compensate for the slower target finding rate and the effect of multiple fast returns due to sliding. It is worthwhile mentioning that, from an evolutionary point of view, changes in dissociation rate could be acquired relatively easily via small mutations in target sequence and/or TF DNA-binding domain \citep{maerkl_2007}. 

We choose the target dissociation rate of the non-sliding TF such that the average mRNA level remains unchanged and, thus, we do not consider a decrease in dissociation rate due to multiple returns of the TF to the binding site as in \cite{zon_2006}. The change in the noise in our model is only due to the accelerated target finding. If we did not correct the dissociation rate, a simple non-sliding lacI equivalent would show both slower target finding rate as well as higher effective target dissociation rate due to the lack of multiple returns. However such a direct comparison would lead to very different average mRNA levels. Using our comparison, we are able to show that the increase in target finding rate due to facilitated diffusion can indeed pose an evolutionary advantage for the cell by decreasing the steady state expression noise of the controlled gene for a specific average expression rate.

\begin{acknowledgments}
We would like to thank Dr Boris Adryan and his group for useful comments and discussions on the manuscript.
\emph{Funding:} This work was supported by the Medical Research Council [G1002110]. A.S. was supported by a BBSRC studentship.
\end{acknowledgments}

\bibliography{facilitated_diffusion_noise.bib}

\begin{thebibliography}{54}%
\makeatletter
\providecommand \@ifxundefined [1]{%
 \@ifx{#1\undefined}
}%
\providecommand \@ifnum [1]{%
 \ifnum #1\expandafter \@firstoftwo
 \else \expandafter \@secondoftwo
 \fi
}%
\providecommand \@ifx [1]{%
 \ifx #1\expandafter \@firstoftwo
 \else \expandafter \@secondoftwo
 \fi
}%
\providecommand \natexlab [1]{#1}%
\providecommand \enquote  [1]{``#1''}%
\providecommand \bibnamefont  [1]{#1}%
\providecommand \bibfnamefont [1]{#1}%
\providecommand \citenamefont [1]{#1}%
\providecommand \href@noop [0]{\@secondoftwo}%
\providecommand \href [0]{\begingroup \@sanitize@url \@href}%
\providecommand \@href[1]{\@@startlink{#1}\@@href}%
\providecommand \@@href[1]{\endgroup#1\@@endlink}%
\providecommand \@sanitize@url [0]{\catcode `\\12\catcode `\$12\catcode
  `\&12\catcode `\#12\catcode `\^12\catcode `\_12\catcode `\%12\relax}%
\providecommand \@@startlink[1]{}%
\providecommand \@@endlink[0]{}%
\providecommand \url  [0]{\begingroup\@sanitize@url \@url }%
\providecommand \@url [1]{\endgroup\@href {#1}{\urlprefix }}%
\providecommand \urlprefix  [0]{URL }%
\providecommand \Eprint [0]{\href }%
\providecommand \doibase [0]{http://dx.doi.org/}%
\providecommand \selectlanguage [0]{\@gobble}%
\providecommand \bibinfo  [0]{\@secondoftwo}%
\providecommand \bibfield  [0]{\@secondoftwo}%
\providecommand \translation [1]{[#1]}%
\providecommand \BibitemOpen [0]{}%
\providecommand \bibitemStop [0]{}%
\providecommand \bibitemNoStop [0]{.\EOS\space}%
\providecommand \EOS [0]{\spacefactor3000\relax}%
\providecommand \BibitemShut  [1]{\csname bibitem#1\endcsname}%
\let\auto@bib@innerbib\@empty
\bibitem [{\citenamefont {Golding}\ \emph {et~al.}(2005)\citenamefont
  {Golding}, \citenamefont {Paulsson}, \citenamefont {Zawilski},\ and\
  \citenamefont {Cox}}]{golding_2005}%
  \BibitemOpen
  \bibfield  {author} {\bibinfo {author} {\bibfnamefont {I.}~\bibnamefont
  {Golding}}, \bibinfo {author} {\bibfnamefont {J.}~\bibnamefont {Paulsson}},
  \bibinfo {author} {\bibfnamefont {S.~M.}\ \bibnamefont {Zawilski}}, \ and\
  \bibinfo {author} {\bibfnamefont {E.~C.}\ \bibnamefont {Cox}},\ }\href
  {\doibase 10.1016/j.cell.2005.09.031} {\bibfield  {journal} {\bibinfo
  {journal} {Cell}\ }\textbf {\bibinfo {volume} {123}},\ \bibinfo {pages}
  {1025} (\bibinfo {year} {2005})}\BibitemShut {NoStop}%
\bibitem [{\citenamefont {Elf}\ \emph {et~al.}(2007)\citenamefont {Elf},
  \citenamefont {Li},\ and\ \citenamefont {Xie}}]{elf_2007}%
  \BibitemOpen
  \bibfield  {author} {\bibinfo {author} {\bibfnamefont {J.}~\bibnamefont
  {Elf}}, \bibinfo {author} {\bibfnamefont {G.-W.}\ \bibnamefont {Li}}, \ and\
  \bibinfo {author} {\bibfnamefont {X.~S.}\ \bibnamefont {Xie}},\ }\href
  {\doibase 10.1126/science.114196} {\bibfield  {journal} {\bibinfo  {journal}
  {Science}\ }\textbf {\bibinfo {volume} {316}},\ \bibinfo {pages} {1191}
  (\bibinfo {year} {2007})}\BibitemShut {NoStop}%
\bibitem [{\citenamefont {Hammar}\ \emph {et~al.}(2012)\citenamefont {Hammar},
  \citenamefont {Leroy}, \citenamefont {Mahmutovic}, \citenamefont {Marklund},
  \citenamefont {Berg},\ and\ \citenamefont {Elf}}]{hammar_2012}%
  \BibitemOpen
  \bibfield  {author} {\bibinfo {author} {\bibfnamefont {P.}~\bibnamefont
  {Hammar}}, \bibinfo {author} {\bibfnamefont {P.}~\bibnamefont {Leroy}},
  \bibinfo {author} {\bibfnamefont {A.}~\bibnamefont {Mahmutovic}}, \bibinfo
  {author} {\bibfnamefont {E.~G.}\ \bibnamefont {Marklund}}, \bibinfo {author}
  {\bibfnamefont {O.~G.}\ \bibnamefont {Berg}}, \ and\ \bibinfo {author}
  {\bibfnamefont {J.}~\bibnamefont {Elf}},\ }\href {\doibase
  10.1126/science.1221648} {\bibfield  {journal} {\bibinfo  {journal}
  {Science}\ }\textbf {\bibinfo {volume} {336}},\ \bibinfo {pages} {1595}
  (\bibinfo {year} {2012})}\BibitemShut {NoStop}%
\bibitem [{\citenamefont {Paulsson}(2004)}]{paulsson_2004}%
  \BibitemOpen
  \bibfield  {author} {\bibinfo {author} {\bibfnamefont {J.}~\bibnamefont
  {Paulsson}},\ }\href {\doibase 10.1038/nature02257} {\bibfield  {journal}
  {\bibinfo  {journal} {Nature}\ }\textbf {\bibinfo {volume} {427}},\ \bibinfo
  {pages} {415} (\bibinfo {year} {2004})}\BibitemShut {NoStop}%
\bibitem [{\citenamefont {Friedman}\ \emph {et~al.}(2006)\citenamefont
  {Friedman}, \citenamefont {Cai},\ and\ \citenamefont {Xie}}]{friedman_2006}%
  \BibitemOpen
  \bibfield  {author} {\bibinfo {author} {\bibfnamefont {N.}~\bibnamefont
  {Friedman}}, \bibinfo {author} {\bibfnamefont {L.}~\bibnamefont {Cai}}, \
  and\ \bibinfo {author} {\bibfnamefont {X.~S.}\ \bibnamefont {Xie}},\ }\href
  {\doibase 10.1103/PhysRevLett.97.168302} {\bibfield  {journal} {\bibinfo
  {journal} {Phys. Rev. Lett.}\ }\textbf {\bibinfo {volume} {97}},\ \bibinfo
  {pages} {168302} (\bibinfo {year} {2006})}\BibitemShut {NoStop}%
\bibitem [{\citenamefont {Mirny}\ \emph {et~al.}(2009)\citenamefont {Mirny},
  \citenamefont {Slutsky}, \citenamefont {Wunderlich}, \citenamefont {Tafvizi},
  \citenamefont {Leith},\ and\ \citenamefont {Kosmrlj}}]{mirny_2009}%
  \BibitemOpen
  \bibfield  {author} {\bibinfo {author} {\bibfnamefont {L.}~\bibnamefont
  {Mirny}}, \bibinfo {author} {\bibfnamefont {M.}~\bibnamefont {Slutsky}},
  \bibinfo {author} {\bibfnamefont {Z.}~\bibnamefont {Wunderlich}}, \bibinfo
  {author} {\bibfnamefont {A.}~\bibnamefont {Tafvizi}}, \bibinfo {author}
  {\bibfnamefont {J.}~\bibnamefont {Leith}}, \ and\ \bibinfo {author}
  {\bibfnamefont {A.}~\bibnamefont {Kosmrlj}},\ }\href {\doibase
  10.1088/1751-8113/42/43/434013} {\bibfield  {journal} {\bibinfo  {journal}
  {J. Phys. A: Math. Theor.}\ }\textbf {\bibinfo {volume} {42}},\ \bibinfo
  {pages} {434013} (\bibinfo {year} {2009})}\BibitemShut {NoStop}%
\bibitem [{\citenamefont {Benichou}\ \emph {et~al.}(2011)\citenamefont
  {Benichou}, \citenamefont {Chevalier}, \citenamefont {Meyer},\ and\
  \citenamefont {Voituriez}}]{benichou_2011}%
  \BibitemOpen
  \bibfield  {author} {\bibinfo {author} {\bibfnamefont {O.}~\bibnamefont
  {Benichou}}, \bibinfo {author} {\bibfnamefont {C.}~\bibnamefont {Chevalier}},
  \bibinfo {author} {\bibfnamefont {B.}~\bibnamefont {Meyer}}, \ and\ \bibinfo
  {author} {\bibfnamefont {R.}~\bibnamefont {Voituriez}},\ }\href {\doibase
  10.1103/PhysRevLett.106.038102} {\bibfield  {journal} {\bibinfo  {journal}
  {Phys. Rev. Lett.}\ }\textbf {\bibinfo {volume} {106}},\ \bibinfo {pages}
  {038102} (\bibinfo {year} {2011})}\BibitemShut {NoStop}%
\bibitem [{\citenamefont {Zabet}\ and\ \citenamefont
  {Adryan}(2012{\natexlab{a}})}]{zabet_2012_model}%
  \BibitemOpen
  \bibfield  {author} {\bibinfo {author} {\bibfnamefont {N.~R.}\ \bibnamefont
  {Zabet}}\ and\ \bibinfo {author} {\bibfnamefont {B.}~\bibnamefont {Adryan}},\
  }\href {\doibase 10.1093/bioinformatics/bts178} {\bibfield  {journal}
  {\bibinfo  {journal} {Bioinformatics}\ }\textbf {\bibinfo {volume} {28}},\
  \bibinfo {pages} {1517} (\bibinfo {year} {2012}{\natexlab{a}})}\BibitemShut
  {NoStop}%
\bibitem [{\citenamefont {Zabet}\ and\ \citenamefont
  {Adryan}(2013)}]{zabet_2013_time}%
  \BibitemOpen
  \bibfield  {author} {\bibinfo {author} {\bibfnamefont {N.~R.}\ \bibnamefont
  {Zabet}}\ and\ \bibinfo {author} {\bibfnamefont {B.}~\bibnamefont {Adryan}},\
  }\href {\doibase 10.3389/fgene.2013.00197} {\bibfield  {journal} {\bibinfo
  {journal} {Front. Genet.}\ }\textbf {\bibinfo {volume} {4}},\ \bibinfo
  {pages} {197} (\bibinfo {year} {2013})}\BibitemShut {NoStop}%
\bibitem [{\citenamefont {Meyer}\ \emph {et~al.}(2012)\citenamefont {Meyer},
  \citenamefont {Benichou}, \citenamefont {Kafri},\ and\ \citenamefont
  {Voituriez}}]{meyer_2012}%
  \BibitemOpen
  \bibfield  {author} {\bibinfo {author} {\bibfnamefont {B.}~\bibnamefont
  {Meyer}}, \bibinfo {author} {\bibfnamefont {O.}~\bibnamefont {Benichou}},
  \bibinfo {author} {\bibfnamefont {Y.}~\bibnamefont {Kafri}}, \ and\ \bibinfo
  {author} {\bibfnamefont {R.}~\bibnamefont {Voituriez}},\ }\href {\doibase
  10.1016/j.bpj.2012.03.060} {\bibfield  {journal} {\bibinfo  {journal}
  {Biophys. J.}\ }\textbf {\bibinfo {volume} {102}},\ \bibinfo {pages} {2186 }
  (\bibinfo {year} {2012})}\BibitemShut {NoStop}%
\bibitem [{\citenamefont {Pulkkinen}\ and\ \citenamefont
  {Metzler}(2013)}]{pulkkinen_2013}%
  \BibitemOpen
  \bibfield  {author} {\bibinfo {author} {\bibfnamefont {O.}~\bibnamefont
  {Pulkkinen}}\ and\ \bibinfo {author} {\bibfnamefont {R.}~\bibnamefont
  {Metzler}},\ }\href {\doibase 10.1103/PhysRevLett.110.198101} {\bibfield
  {journal} {\bibinfo  {journal} {Phys. Rev. Lett.}\ }\textbf {\bibinfo
  {volume} {110}},\ \bibinfo {pages} {198101} (\bibinfo {year}
  {2013})}\BibitemShut {NoStop}%
\bibitem [{\citenamefont {Sharon}\ \emph {et~al.}(2014)\citenamefont {Sharon},
  \citenamefont {van Dijk}, \citenamefont {Kalma}, \citenamefont {Keren},
  \citenamefont {Yakhini},\ and\ \citenamefont {Segal}}]{sharon_2014}%
  \BibitemOpen
  \bibfield  {author} {\bibinfo {author} {\bibfnamefont {E.}~\bibnamefont
  {Sharon}}, \bibinfo {author} {\bibfnamefont {D.}~\bibnamefont {van Dijk}},
  \bibinfo {author} {\bibfnamefont {Y.}~\bibnamefont {Kalma}}, \bibinfo
  {author} {\bibfnamefont {L.}~\bibnamefont {Keren}}, \bibinfo {author}
  {\bibfnamefont {O.~M.~Z.}\ \bibnamefont {Yakhini}}, \ and\ \bibinfo {author}
  {\bibfnamefont {E.}~\bibnamefont {Segal}},\ }\href {\doibase
  10.1101/gr.168773.113} {\bibfield  {journal} {\bibinfo  {journal} {Genome
  Resarch}\ } (\bibinfo {year} {2014}),\ 10.1101/gr.168773.113}\BibitemShut
  {NoStop}%
\bibitem [{\citenamefont {Riggs}\ \emph {et~al.}(1970)\citenamefont {Riggs},
  \citenamefont {Bourgeois},\ and\ \citenamefont {Cohn}}]{riggs_1970a}%
  \BibitemOpen
  \bibfield  {author} {\bibinfo {author} {\bibfnamefont {A.~D.}\ \bibnamefont
  {Riggs}}, \bibinfo {author} {\bibfnamefont {S.}~\bibnamefont {Bourgeois}}, \
  and\ \bibinfo {author} {\bibfnamefont {M.}~\bibnamefont {Cohn}},\ }\href
  {\doibase 10.1016/0022-2836(70)90074-4} {\bibfield  {journal} {\bibinfo
  {journal} {J. Mol. Biol.}\ }\textbf {\bibinfo {volume} {53}},\ \bibinfo
  {pages} {401} (\bibinfo {year} {1970})}\BibitemShut {NoStop}%
\bibitem [{\citenamefont {Berg}\ \emph {et~al.}(1981)\citenamefont {Berg},
  \citenamefont {Winter},\ and\ \citenamefont {von Hippel}}]{berg_1981}%
  \BibitemOpen
  \bibfield  {author} {\bibinfo {author} {\bibfnamefont {O.~G.}\ \bibnamefont
  {Berg}}, \bibinfo {author} {\bibfnamefont {R.~B.}\ \bibnamefont {Winter}}, \
  and\ \bibinfo {author} {\bibfnamefont {P.~H.}\ \bibnamefont {von Hippel}},\
  }\href {\doibase 10.1021/bi00527a028} {\bibfield  {journal} {\bibinfo
  {journal} {Biochemistry}\ }\textbf {\bibinfo {volume} {20}},\ \bibinfo
  {pages} {6929} (\bibinfo {year} {1981})}\BibitemShut {NoStop}%
\bibitem [{Note1()}]{Note1}%
  \BibitemOpen
  \bibinfo {note} {Since it is difficult to experimentally distinguish between
  the 1D translocation modes, by sliding we refer to 1D random walk which
  includes both sliding and hopping}\BibitemShut {NoStop}%
\bibitem [{\citenamefont {Kabata}\ \emph {et~al.}(1993)\citenamefont {Kabata},
  \citenamefont {Kurosawa}, \citenamefont {I~Arai}, \citenamefont {Margarson},
  \citenamefont {Glass},\ and\ \citenamefont {Shimamoto}}]{kabata_1993}%
  \BibitemOpen
  \bibfield  {author} {\bibinfo {author} {\bibfnamefont {H.}~\bibnamefont
  {Kabata}}, \bibinfo {author} {\bibfnamefont {O.}~\bibnamefont {Kurosawa}},
  \bibinfo {author} {\bibfnamefont {M.~W.}\ \bibnamefont {I~Arai}}, \bibinfo
  {author} {\bibfnamefont {S.}~\bibnamefont {Margarson}}, \bibinfo {author}
  {\bibfnamefont {R.~E.}\ \bibnamefont {Glass}}, \ and\ \bibinfo {author}
  {\bibfnamefont {N.}~\bibnamefont {Shimamoto}},\ }\href {\doibase
  10.1126/science.8248804} {\bibfield  {journal} {\bibinfo  {journal}
  {Science}\ }\textbf {\bibinfo {volume} {262}},\ \bibinfo {pages} {1561}
  (\bibinfo {year} {1993})}\BibitemShut {NoStop}%
\bibitem [{\citenamefont {Redner}(2001)}]{redner_2001}%
  \BibitemOpen
  \bibfield  {author} {\bibinfo {author} {\bibfnamefont {S.}~\bibnamefont
  {Redner}},\ }\href@noop {} {\emph {\bibinfo {title} {A Guide to First-Passage
  Processes}}}\ (\bibinfo  {publisher} {Cambridge University Press},\ \bibinfo
  {address} {New York},\ \bibinfo {year} {2001})\BibitemShut {NoStop}%
\bibitem [{\citenamefont {van Zon}\ \emph {et~al.}(2006)\citenamefont {van
  Zon}, \citenamefont {Morelli}, \citenamefont {Tanase-Nicola},\ and\
  \citenamefont {ten Wolde}}]{zon_2006}%
  \BibitemOpen
  \bibfield  {author} {\bibinfo {author} {\bibfnamefont {J.~S.}\ \bibnamefont
  {van Zon}}, \bibinfo {author} {\bibfnamefont {M.~J.}\ \bibnamefont
  {Morelli}}, \bibinfo {author} {\bibfnamefont {S.}~\bibnamefont
  {Tanase-Nicola}}, \ and\ \bibinfo {author} {\bibfnamefont {P.~R.}\
  \bibnamefont {ten Wolde}},\ }\href {\doibase 10.1529/biophysj.106.086157}
  {\bibfield  {journal} {\bibinfo  {journal} {Biophys. J.}\ }\textbf {\bibinfo
  {volume} {91}},\ \bibinfo {pages} {4350} (\bibinfo {year}
  {2006})}\BibitemShut {NoStop}%
\bibitem [{\citenamefont {Zabet}\ and\ \citenamefont
  {Adryan}(2012{\natexlab{b}})}]{zabet_2012_review}%
  \BibitemOpen
  \bibfield  {author} {\bibinfo {author} {\bibfnamefont {N.~R.}\ \bibnamefont
  {Zabet}}\ and\ \bibinfo {author} {\bibfnamefont {B.}~\bibnamefont {Adryan}},\
  }\href {\doibase 10.1039/C2MB25201E} {\bibfield  {journal} {\bibinfo
  {journal} {Mol. Biosyst.}\ }\textbf {\bibinfo {volume} {8}},\ \bibinfo
  {pages} {2815} (\bibinfo {year} {2012}{\natexlab{b}})}\BibitemShut {NoStop}%
\bibitem [{\citenamefont {Ezer}\ \emph {et~al.}(2014)\citenamefont {Ezer},
  \citenamefont {Zabet},\ and\ \citenamefont {Adryan}}]{ezer_2014}%
  \BibitemOpen
  \bibfield  {author} {\bibinfo {author} {\bibfnamefont {D.}~\bibnamefont
  {Ezer}}, \bibinfo {author} {\bibfnamefont {N.~R.}\ \bibnamefont {Zabet}}, \
  and\ \bibinfo {author} {\bibfnamefont {B.}~\bibnamefont {Adryan}},\ }\href
  {\doibase 10.1093/nar/gku078} {\bibfield  {journal} {\bibinfo  {journal}
  {Nucleic Acids Res.}\ }\textbf {\bibinfo {volume} {42}},\ \bibinfo {pages}
  {4196} (\bibinfo {year} {2014})}\BibitemShut {NoStop}%
\bibitem [{\citenamefont {Wunderlich}\ and\ \citenamefont
  {Mirny}(2008)}]{wunderlich_2008}%
  \BibitemOpen
  \bibfield  {author} {\bibinfo {author} {\bibfnamefont {Z.}~\bibnamefont
  {Wunderlich}}\ and\ \bibinfo {author} {\bibfnamefont {L.~A.}\ \bibnamefont
  {Mirny}},\ }\href {\doibase 10.1093/nar/gkn173} {\bibfield  {journal}
  {\bibinfo  {journal} {Nucleic Acids Res.}\ }\textbf {\bibinfo {volume}
  {36}},\ \bibinfo {pages} {3570} (\bibinfo {year} {2008})}\BibitemShut
  {NoStop}%
\bibitem [{\citenamefont {Gillespie}(1977)}]{gillespie_1977}%
  \BibitemOpen
  \bibfield  {author} {\bibinfo {author} {\bibfnamefont {D.~T.}\ \bibnamefont
  {Gillespie}},\ }\href {\doibase 10.1021/j100540a008} {\bibfield  {journal}
  {\bibinfo  {journal} {J. Phys. Chem.}\ }\textbf {\bibinfo {volume} {81}},\
  \bibinfo {pages} {2340} (\bibinfo {year} {1977})}\BibitemShut {NoStop}%
\bibitem [{\citenamefont {So}\ \emph {et~al.}(2011)\citenamefont {So},
  \citenamefont {Ghosh}, \citenamefont {Zong}, \citenamefont {Sepulveda},
  \citenamefont {Segev},\ and\ \citenamefont {Golding}}]{so_2011}%
  \BibitemOpen
  \bibfield  {author} {\bibinfo {author} {\bibfnamefont {L.-h.}\ \bibnamefont
  {So}}, \bibinfo {author} {\bibfnamefont {A.}~\bibnamefont {Ghosh}}, \bibinfo
  {author} {\bibfnamefont {C.}~\bibnamefont {Zong}}, \bibinfo {author}
  {\bibfnamefont {L.~A.}\ \bibnamefont {Sepulveda}}, \bibinfo {author}
  {\bibfnamefont {R.}~\bibnamefont {Segev}}, \ and\ \bibinfo {author}
  {\bibfnamefont {I.}~\bibnamefont {Golding}},\ }\href {\doibase
  10.1038/ng.821} {\bibfield  {journal} {\bibinfo  {journal} {Nat. Genet.}\
  }\textbf {\bibinfo {volume} {43}},\ \bibinfo {pages} {554} (\bibinfo {year}
  {2011})}\BibitemShut {NoStop}%
\bibitem [{\citenamefont {Chu}\ \emph {et~al.}(2009)\citenamefont {Chu},
  \citenamefont {Zabet},\ and\ \citenamefont {Mitavskiy}}]{chu_2009}%
  \BibitemOpen
  \bibfield  {author} {\bibinfo {author} {\bibfnamefont {D.}~\bibnamefont
  {Chu}}, \bibinfo {author} {\bibfnamefont {N.~R.}\ \bibnamefont {Zabet}}, \
  and\ \bibinfo {author} {\bibfnamefont {B.}~\bibnamefont {Mitavskiy}},\ }\href
  {\doibase 10.1016/j.jtbi.2008.11.026} {\bibfield  {journal} {\bibinfo
  {journal} {J. Theor. Biol.}\ }\textbf {\bibinfo {volume} {257}},\ \bibinfo
  {pages} {419} (\bibinfo {year} {2009})}\BibitemShut {NoStop}%
\bibitem [{\citenamefont {Paulsson}(2005)}]{paulsson_2005}%
  \BibitemOpen
  \bibfield  {author} {\bibinfo {author} {\bibfnamefont {J.}~\bibnamefont
  {Paulsson}},\ }\href {\doibase 10.1016/j.plrev.2005.03.003} {\bibfield
  {journal} {\bibinfo  {journal} {Phys. Life Rev.}\ }\textbf {\bibinfo {volume}
  {2}},\ \bibinfo {pages} {157} (\bibinfo {year} {2005})}\BibitemShut {NoStop}%
\bibitem [{\citenamefont {Slutsky}\ and\ \citenamefont
  {Mirny}(2004)}]{slutsky_2004b}%
  \BibitemOpen
  \bibfield  {author} {\bibinfo {author} {\bibfnamefont {M.}~\bibnamefont
  {Slutsky}}\ and\ \bibinfo {author} {\bibfnamefont {L.~A.}\ \bibnamefont
  {Mirny}},\ }\href {\doibase 10.1529/biophysj.104.050765} {\bibfield
  {journal} {\bibinfo  {journal} {Biophys. J.}\ }\textbf {\bibinfo {volume}
  {87}},\ \bibinfo {pages} {4021} (\bibinfo {year} {2004})}\BibitemShut
  {NoStop}%
\bibitem [{\citenamefont {Bar-Even}\ \emph {et~al.}(2006)\citenamefont
  {Bar-Even}, \citenamefont {Paulsson}, \citenamefont {Maheshri}, \citenamefont
  {Carmi}, \citenamefont {O'Shea}, \citenamefont {Pilpel},\ and\ \citenamefont
  {Barkai}}]{even_2006}%
  \BibitemOpen
  \bibfield  {author} {\bibinfo {author} {\bibfnamefont {A.}~\bibnamefont
  {Bar-Even}}, \bibinfo {author} {\bibfnamefont {J.}~\bibnamefont {Paulsson}},
  \bibinfo {author} {\bibfnamefont {N.}~\bibnamefont {Maheshri}}, \bibinfo
  {author} {\bibfnamefont {M.}~\bibnamefont {Carmi}}, \bibinfo {author}
  {\bibfnamefont {E.}~\bibnamefont {O'Shea}}, \bibinfo {author} {\bibfnamefont
  {Y.}~\bibnamefont {Pilpel}}, \ and\ \bibinfo {author} {\bibfnamefont
  {N.}~\bibnamefont {Barkai}},\ }\href {\doibase 10.1038/ng1807} {\bibfield
  {journal} {\bibinfo  {journal} {Nat. Genet.}\ }\textbf {\bibinfo {volume}
  {38}},\ \bibinfo {pages} {636} (\bibinfo {year} {2006})}\BibitemShut
  {NoStop}%
\bibitem [{\citenamefont {Newman}\ \emph {et~al.}(2006)\citenamefont {Newman},
  \citenamefont {Ghaemmaghami}, \citenamefont {Ihmels}, \citenamefont
  {Breslow}, \citenamefont {Noble}, \citenamefont {DeRisi},\ and\ \citenamefont
  {Weissman}}]{newman_2006}%
  \BibitemOpen
  \bibfield  {author} {\bibinfo {author} {\bibfnamefont {J.~R.~S.}\
  \bibnamefont {Newman}}, \bibinfo {author} {\bibfnamefont {S.}~\bibnamefont
  {Ghaemmaghami}}, \bibinfo {author} {\bibfnamefont {J.}~\bibnamefont
  {Ihmels}}, \bibinfo {author} {\bibfnamefont {D.~K.}\ \bibnamefont {Breslow}},
  \bibinfo {author} {\bibfnamefont {M.}~\bibnamefont {Noble}}, \bibinfo
  {author} {\bibfnamefont {J.~L.}\ \bibnamefont {DeRisi}}, \ and\ \bibinfo
  {author} {\bibfnamefont {J.~S.}\ \bibnamefont {Weissman}},\ }\href {\doibase
  10.1038/nature04785} {\bibfield  {journal} {\bibinfo  {journal} {Nature}\
  }\textbf {\bibinfo {volume} {441}},\ \bibinfo {pages} {840} (\bibinfo {year}
  {2006})}\BibitemShut {NoStop}%
\bibitem [{\citenamefont {Taniguchi}\ \emph {et~al.}(2010)\citenamefont
  {Taniguchi}, \citenamefont {Choi}, \citenamefont {Li}, \citenamefont {Chen},
  \citenamefont {Babu}, \citenamefont {Hearn}, \citenamefont {Emili},\ and\
  \citenamefont {Xie}}]{taniguchi_2010}%
  \BibitemOpen
  \bibfield  {author} {\bibinfo {author} {\bibfnamefont {Y.}~\bibnamefont
  {Taniguchi}}, \bibinfo {author} {\bibfnamefont {P.~J.}\ \bibnamefont {Choi}},
  \bibinfo {author} {\bibfnamefont {G.-W.}\ \bibnamefont {Li}}, \bibinfo
  {author} {\bibfnamefont {H.}~\bibnamefont {Chen}}, \bibinfo {author}
  {\bibfnamefont {M.}~\bibnamefont {Babu}}, \bibinfo {author} {\bibfnamefont
  {J.}~\bibnamefont {Hearn}}, \bibinfo {author} {\bibfnamefont
  {A.}~\bibnamefont {Emili}}, \ and\ \bibinfo {author} {\bibfnamefont {X.~S.}\
  \bibnamefont {Xie}},\ }\href {\doibase 10.1126/science.1188308} {\bibfield
  {journal} {\bibinfo  {journal} {Science}\ }\textbf {\bibinfo {volume}
  {329}},\ \bibinfo {pages} {533} (\bibinfo {year} {2010})}\BibitemShut
  {NoStop}%
\bibitem [{\citenamefont {DeSantis}\ \emph {et~al.}(2011)\citenamefont
  {DeSantis}, \citenamefont {Li},\ and\ \citenamefont {Wang}}]{desantis_2011}%
  \BibitemOpen
  \bibfield  {author} {\bibinfo {author} {\bibfnamefont {M.~C.}\ \bibnamefont
  {DeSantis}}, \bibinfo {author} {\bibfnamefont {J.-L.}\ \bibnamefont {Li}}, \
  and\ \bibinfo {author} {\bibfnamefont {Y.~M.}\ \bibnamefont {Wang}},\ }\href
  {\doibase 10.1103/PhysRevE.83.021907} {\bibfield  {journal} {\bibinfo
  {journal} {Phys. Rev. E}\ }\textbf {\bibinfo {volume} {83}},\ \bibinfo
  {pages} {021907} (\bibinfo {year} {2011})}\BibitemShut {NoStop}%
\bibitem [{\citenamefont {Sanchez}\ and\ \citenamefont
  {Golding}(2013)}]{sanchez_2013}%
  \BibitemOpen
  \bibfield  {author} {\bibinfo {author} {\bibfnamefont {A.}~\bibnamefont
  {Sanchez}}\ and\ \bibinfo {author} {\bibfnamefont {I.}~\bibnamefont
  {Golding}},\ }\href {\doibase 10.1126/science.1242975} {\bibfield  {journal}
  {\bibinfo  {journal} {Science}\ }\textbf {\bibinfo {volume} {342}},\ \bibinfo
  {pages} {1188} (\bibinfo {year} {2013})}\BibitemShut {NoStop}%
\bibitem [{\citenamefont {Bothma}\ \emph {et~al.}(2014)\citenamefont {Bothma},
  \citenamefont {Garcia}, \citenamefont {Esposito}, \citenamefont {Schlissel},
  \citenamefont {Gregor},\ and\ \citenamefont {Levine}}]{bothma_2014}%
  \BibitemOpen
  \bibfield  {author} {\bibinfo {author} {\bibfnamefont {J.~P.}\ \bibnamefont
  {Bothma}}, \bibinfo {author} {\bibfnamefont {H.~G.}\ \bibnamefont {Garcia}},
  \bibinfo {author} {\bibfnamefont {E.}~\bibnamefont {Esposito}}, \bibinfo
  {author} {\bibfnamefont {G.}~\bibnamefont {Schlissel}}, \bibinfo {author}
  {\bibfnamefont {T.}~\bibnamefont {Gregor}}, \ and\ \bibinfo {author}
  {\bibfnamefont {M.}~\bibnamefont {Levine}},\ }\href {\doibase
  10.1073/pnas.1410022111} {\bibfield  {journal} {\bibinfo  {journal} {Proc.
  Natl. Acad. Sci.}\ }\textbf {\bibinfo {volume} {111}},\ \bibinfo {pages}
  {10598} (\bibinfo {year} {2014})}\BibitemShut {NoStop}%
\bibitem [{\citenamefont {Zabet}\ and\ \citenamefont {Chu}(2010)}]{zabet_2009}%
  \BibitemOpen
  \bibfield  {author} {\bibinfo {author} {\bibfnamefont {N.~R.}\ \bibnamefont
  {Zabet}}\ and\ \bibinfo {author} {\bibfnamefont {D.~F.}\ \bibnamefont
  {Chu}},\ }\href {\doibase 10.1098/rsif.2009.0474} {\bibfield  {journal}
  {\bibinfo  {journal} {J. R. Soc. Interface}\ }\textbf {\bibinfo {volume}
  {7}},\ \bibinfo {pages} {945} (\bibinfo {year} {2010})}\BibitemShut {NoStop}%
\bibitem [{\citenamefont {Zabet}(2011)}]{zabet_2011}%
  \BibitemOpen
  \bibfield  {author} {\bibinfo {author} {\bibfnamefont {N.~R.}\ \bibnamefont
  {Zabet}},\ }\href {\doibase 10.1016/j.jtbi.2011.06.021} {\bibfield  {journal}
  {\bibinfo  {journal} {J. Theor. Biol.}\ }\textbf {\bibinfo {volume} {284}},\
  \bibinfo {pages} {82} (\bibinfo {year} {2011})}\BibitemShut {NoStop}%
\bibitem [{\citenamefont {Brackley}\ \emph {et~al.}(2013)\citenamefont
  {Brackley}, \citenamefont {Cates},\ and\ \citenamefont
  {Marenduzzo}}]{brackley_2013_crowding}%
  \BibitemOpen
  \bibfield  {author} {\bibinfo {author} {\bibfnamefont {C.~A.}\ \bibnamefont
  {Brackley}}, \bibinfo {author} {\bibfnamefont {M.~E.}\ \bibnamefont {Cates}},
  \ and\ \bibinfo {author} {\bibfnamefont {D.}~\bibnamefont {Marenduzzo}},\
  }\href {\doibase 10.1103/PhysRevLett.111.108101} {\bibfield  {journal}
  {\bibinfo  {journal} {Phys. Rev. Lett.}\ }\textbf {\bibinfo {volume} {111}},\
  \bibinfo {pages} {108101} (\bibinfo {year} {2013})}\BibitemShut {NoStop}%
\bibitem [{\citenamefont {Vuzman}\ and\ \citenamefont
  {Levy}(2010)}]{vuzman_2010_tails_composition}%
  \BibitemOpen
  \bibfield  {author} {\bibinfo {author} {\bibfnamefont {D.}~\bibnamefont
  {Vuzman}}\ and\ \bibinfo {author} {\bibfnamefont {Y.}~\bibnamefont {Levy}},\
  }\href {\doibase 10.1073/pnas.1011775107} {\bibfield  {journal} {\bibinfo
  {journal} {Proc. Natl. Acad. Sci.}\ }\textbf {\bibinfo {volume} {107}},\
  \bibinfo {pages} {21004} (\bibinfo {year} {2010})}\BibitemShut {NoStop}%
\bibitem [{\citenamefont {Tafvizi}\ \emph {et~al.}(2011)\citenamefont
  {Tafvizi}, \citenamefont {Huang}, \citenamefont {Fersht}, \citenamefont
  {Mirny},\ and\ \citenamefont {van Oijen}}]{tafvizi_2011}%
  \BibitemOpen
  \bibfield  {author} {\bibinfo {author} {\bibfnamefont {A.}~\bibnamefont
  {Tafvizi}}, \bibinfo {author} {\bibfnamefont {F.}~\bibnamefont {Huang}},
  \bibinfo {author} {\bibfnamefont {A.~R.}\ \bibnamefont {Fersht}}, \bibinfo
  {author} {\bibfnamefont {L.~A.}\ \bibnamefont {Mirny}}, \ and\ \bibinfo
  {author} {\bibfnamefont {A.~M.}\ \bibnamefont {van Oijen}},\ }\href {\doibase
  10.1073/pnas.1016020107} {\bibfield  {journal} {\bibinfo  {journal} {Proc.
  Natl. Acad. Sci.}\ }\textbf {\bibinfo {volume} {108}},\ \bibinfo {pages}
  {563} (\bibinfo {year} {2011})}\BibitemShut {NoStop}%
\bibitem [{\citenamefont {Klafter}\ and\ \citenamefont
  {Sokolov}(2011)}]{klafter_2011}%
  \BibitemOpen
  \bibfield  {author} {\bibinfo {author} {\bibfnamefont {J.}~\bibnamefont
  {Klafter}}\ and\ \bibinfo {author} {\bibfnamefont {I.~M.}\ \bibnamefont
  {Sokolov}},\ }\href@noop {} {\emph {\bibinfo {title} {First Steps in Random
  Walks, From Tools to Applications}}}\ (\bibinfo  {publisher} {Oxford Univ
  Press},\ \bibinfo {address} {New York},\ \bibinfo {year} {2011})\BibitemShut
  {NoStop}%
\bibitem [{\citenamefont {Gilbert}\ and\ \citenamefont
  {M{\"u}ller-Hill}(1966)}]{gilbert_1966}%
  \BibitemOpen
  \bibfield  {author} {\bibinfo {author} {\bibfnamefont {W.}~\bibnamefont
  {Gilbert}}\ and\ \bibinfo {author} {\bibfnamefont {B.}~\bibnamefont
  {M{\"u}ller-Hill}},\ }\href@noop {} {\bibfield  {journal} {\bibinfo
  {journal} {Proc. Natl. Acad. Sci.}\ }\textbf {\bibinfo {volume} {56}},\
  \bibinfo {pages} {1891} (\bibinfo {year} {1966})}\BibitemShut {NoStop}%
\bibitem [{\citenamefont {Li}\ \emph {et~al.}(2009)\citenamefont {Li},
  \citenamefont {Berg},\ and\ \citenamefont {Elf}}]{li_2009}%
  \BibitemOpen
  \bibfield  {author} {\bibinfo {author} {\bibfnamefont {G.-W.}\ \bibnamefont
  {Li}}, \bibinfo {author} {\bibfnamefont {O.~G.}\ \bibnamefont {Berg}}, \ and\
  \bibinfo {author} {\bibfnamefont {J.}~\bibnamefont {Elf}},\ }\href {\doibase
  10.1038/NPHYS1222} {\bibfield  {journal} {\bibinfo  {journal} {Nat. Phys.}\
  }\textbf {\bibinfo {volume} {5}},\ \bibinfo {pages} {294} (\bibinfo {year}
  {2009})}\BibitemShut {NoStop}%
\bibitem [{\citenamefont {Fried}\ and\ \citenamefont
  {Crothers}(1981)}]{fried_1981}%
  \BibitemOpen
  \bibfield  {author} {\bibinfo {author} {\bibfnamefont {M.}~\bibnamefont
  {Fried}}\ and\ \bibinfo {author} {\bibfnamefont {D.~M.}\ \bibnamefont
  {Crothers}},\ }\href {\doibase 10.1093/nar/9.23.6505} {\bibfield  {journal}
  {\bibinfo  {journal} {Nucleic Acids Res.}\ }\textbf {\bibinfo {volume} {9}},\
  \bibinfo {pages} {6505} (\bibinfo {year} {1981})}\BibitemShut {NoStop}%
\bibitem [{\citenamefont {Malan}\ \emph {et~al.}(1984)\citenamefont {Malan},
  \citenamefont {Kolb}, \citenamefont {Buc},\ and\ \citenamefont
  {McClure}}]{malan_1984}%
  \BibitemOpen
  \bibfield  {author} {\bibinfo {author} {\bibfnamefont {T.}~\bibnamefont
  {Malan}}, \bibinfo {author} {\bibfnamefont {A.}~\bibnamefont {Kolb}},
  \bibinfo {author} {\bibfnamefont {H.}~\bibnamefont {Buc}}, \ and\ \bibinfo
  {author} {\bibfnamefont {W.~R.}\ \bibnamefont {McClure}},\ }\href {\doibase
  http://dx.doi.org/10.1016/0022-2836(84)90262-6} {\bibfield  {journal}
  {\bibinfo  {journal} {J. Mol. Biol.}\ }\textbf {\bibinfo {volume} {180}},\
  \bibinfo {pages} {881} (\bibinfo {year} {1984})}\BibitemShut {NoStop}%
\bibitem [{\citenamefont {Kennell}\ and\ \citenamefont
  {Riezman}(1977)}]{kennell_1977}%
  \BibitemOpen
  \bibfield  {author} {\bibinfo {author} {\bibfnamefont {D.}~\bibnamefont
  {Kennell}}\ and\ \bibinfo {author} {\bibfnamefont {H.}~\bibnamefont
  {Riezman}},\ }\href {\doibase http://dx.doi.org/10.1016/0022-2836(77)90279-0}
  {\bibfield  {journal} {\bibinfo  {journal} {J. Mol. Biol.}\ }\textbf
  {\bibinfo {volume} {114}},\ \bibinfo {pages} {1} (\bibinfo {year}
  {1977})}\BibitemShut {NoStop}%
\bibitem [{\citenamefont {Selinger}\ \emph {et~al.}(2003)\citenamefont
  {Selinger}, \citenamefont {Saxena}, \citenamefont {Cheung}, \citenamefont
  {Church},\ and\ \citenamefont {Rosenow}}]{selinger_2003}%
  \BibitemOpen
  \bibfield  {author} {\bibinfo {author} {\bibfnamefont {D.~W.}\ \bibnamefont
  {Selinger}}, \bibinfo {author} {\bibfnamefont {R.~M.}\ \bibnamefont
  {Saxena}}, \bibinfo {author} {\bibfnamefont {K.~J.}\ \bibnamefont {Cheung}},
  \bibinfo {author} {\bibfnamefont {G.~M.}\ \bibnamefont {Church}}, \ and\
  \bibinfo {author} {\bibfnamefont {C.}~\bibnamefont {Rosenow}},\ }\href@noop
  {} {\bibfield  {journal} {\bibinfo  {journal} {Genome Res.}\ }\textbf
  {\bibinfo {volume} {13}},\ \bibinfo {pages} {216} (\bibinfo {year}
  {2003})}\BibitemShut {NoStop}%
\bibitem [{\citenamefont {Ehretsmann}\ \emph {et~al.}(1992)\citenamefont
  {Ehretsmann}, \citenamefont {Carpousis},\ and\ \citenamefont
  {Krisch}}]{ehretsmann_1992}%
  \BibitemOpen
  \bibfield  {author} {\bibinfo {author} {\bibfnamefont {C.~P.}\ \bibnamefont
  {Ehretsmann}}, \bibinfo {author} {\bibfnamefont {A.~J.}\ \bibnamefont
  {Carpousis}}, \ and\ \bibinfo {author} {\bibfnamefont {H.~M.}\ \bibnamefont
  {Krisch}},\ }\href@noop {} {\bibfield  {journal} {\bibinfo  {journal} {FASEB
  J.}\ }\textbf {\bibinfo {volume} {6}},\ \bibinfo {pages} {3186} (\bibinfo
  {year} {1992})}\BibitemShut {NoStop}%
\bibitem [{\citenamefont {Mandelstam}(1957)}]{mandelstam_1957}%
  \BibitemOpen
  \bibfield  {author} {\bibinfo {author} {\bibfnamefont {J.}~\bibnamefont
  {Mandelstam}},\ }\href@noop {} {\bibfield  {journal} {\bibinfo  {journal}
  {Nature}\ }\textbf {\bibinfo {volume} {179}},\ \bibinfo {pages} {1179}
  (\bibinfo {year} {1957})}\BibitemShut {NoStop}%
\bibitem [{\citenamefont {Riley}\ \emph {et~al.}(2006)\citenamefont {Riley},
  \citenamefont {Abe}, \citenamefont {Arnaud}, \citenamefont {Berlyn},
  \citenamefont {Blattner}, \citenamefont {Chaudhuri}, \citenamefont {Glasner},
  \citenamefont {Horiuchi}, \citenamefont {Keseler}, \citenamefont {Kosuge},
  \citenamefont {Mori}, \citenamefont {Perna}, \citenamefont {Plunkett},
  \citenamefont {Rudd}, \citenamefont {Serres}, \citenamefont {Thomas},
  \citenamefont {Thomson}, \citenamefont {Wishart},\ and\ \citenamefont
  {Wanner}}]{riley_2006}%
  \BibitemOpen
  \bibfield  {author} {\bibinfo {author} {\bibfnamefont {M.}~\bibnamefont
  {Riley}}, \bibinfo {author} {\bibfnamefont {T.}~\bibnamefont {Abe}}, \bibinfo
  {author} {\bibfnamefont {M.~B.}\ \bibnamefont {Arnaud}}, \bibinfo {author}
  {\bibfnamefont {M.~K.}\ \bibnamefont {Berlyn}}, \bibinfo {author}
  {\bibfnamefont {F.~R.}\ \bibnamefont {Blattner}}, \bibinfo {author}
  {\bibfnamefont {R.~R.}\ \bibnamefont {Chaudhuri}}, \bibinfo {author}
  {\bibfnamefont {J.~D.}\ \bibnamefont {Glasner}}, \bibinfo {author}
  {\bibfnamefont {T.}~\bibnamefont {Horiuchi}}, \bibinfo {author}
  {\bibfnamefont {I.~M.}\ \bibnamefont {Keseler}}, \bibinfo {author}
  {\bibfnamefont {T.}~\bibnamefont {Kosuge}}, \bibinfo {author} {\bibfnamefont
  {H.}~\bibnamefont {Mori}}, \bibinfo {author} {\bibfnamefont {N.~T.}\
  \bibnamefont {Perna}}, \bibinfo {author} {\bibfnamefont {G.}~\bibnamefont
  {Plunkett}}, \bibinfo {author} {\bibfnamefont {K.~E.}\ \bibnamefont {Rudd}},
  \bibinfo {author} {\bibfnamefont {M.~H.}\ \bibnamefont {Serres}}, \bibinfo
  {author} {\bibfnamefont {G.~H.}\ \bibnamefont {Thomas}}, \bibinfo {author}
  {\bibfnamefont {N.~R.}\ \bibnamefont {Thomson}}, \bibinfo {author}
  {\bibfnamefont {D.}~\bibnamefont {Wishart}}, \ and\ \bibinfo {author}
  {\bibfnamefont {B.~L.}\ \bibnamefont {Wanner}},\ }\href {\doibase
  10.1093/nar/gkj405} {\bibfield  {journal} {\bibinfo  {journal} {Nucleic Acids
  Res.}\ }\textbf {\bibinfo {volume} {34}},\ \bibinfo {pages} {1} (\bibinfo
  {year} {2006})}\BibitemShut {NoStop}%
\bibitem [{\citenamefont {Hammar}\ \emph {et~al.}(2014)\citenamefont {Hammar},
  \citenamefont {Walld\'{e}n}, \citenamefont {Fange}, \citenamefont {Persson},
  \citenamefont {\"{O}zden Baltekin}, \citenamefont {Ullman}, \citenamefont
  {Leroy},\ and\ \citenamefont {Elf}}]{hammar_2014}%
  \BibitemOpen
  \bibfield  {author} {\bibinfo {author} {\bibfnamefont {P.}~\bibnamefont
  {Hammar}}, \bibinfo {author} {\bibfnamefont {M.}~\bibnamefont {Walld\'{e}n}},
  \bibinfo {author} {\bibfnamefont {D.}~\bibnamefont {Fange}}, \bibinfo
  {author} {\bibfnamefont {F.}~\bibnamefont {Persson}}, \bibinfo {author}
  {\bibnamefont {\"{O}zden Baltekin}}, \bibinfo {author} {\bibfnamefont
  {G.}~\bibnamefont {Ullman}}, \bibinfo {author} {\bibfnamefont
  {P.}~\bibnamefont {Leroy}}, \ and\ \bibinfo {author} {\bibfnamefont
  {J.}~\bibnamefont {Elf}},\ }\href {\doibase 10.1038/ng.2905} {\bibfield
  {journal} {\bibinfo  {journal} {Nat. Genet.}\ }\textbf {\bibinfo {volume}
  {46}},\ \bibinfo {pages} {405} (\bibinfo {year} {2014})}\BibitemShut
  {NoStop}%
\bibitem [{\citenamefont {Bremer}\ and\ \citenamefont
  {Dennis}(1996)}]{bremer_1996}%
  \BibitemOpen
  \bibfield  {author} {\bibinfo {author} {\bibfnamefont {H.}~\bibnamefont
  {Bremer}}\ and\ \bibinfo {author} {\bibfnamefont {P.~P.}\ \bibnamefont
  {Dennis}},\ }in\ \href@noop {} {\emph {\bibinfo {booktitle}
  {\emph{Escherichia coli} and \emph{Salmonella}: cellular and molecular
  biology}}},\ \bibinfo {editor} {edited by\ \bibinfo {editor} {\bibfnamefont
  {F.}~\bibnamefont {Neidhardt}}}\ (\bibinfo  {publisher} {ASM Press},\
  \bibinfo {address} {Washington D.C.},\ \bibinfo {year} {1996})\ \bibinfo
  {edition} {2nd}\ ed.,\ pp.\ \bibinfo {pages} {1553--1569.}\BibitemShut
  {Stop}%
\bibitem [{\citenamefont {Hager}\ \emph {et~al.}(2009)\citenamefont {Hager},
  \citenamefont {McNally},\ and\ \citenamefont {Misteli}}]{hager_2009}%
  \BibitemOpen
  \bibfield  {author} {\bibinfo {author} {\bibfnamefont {G.~L.}\ \bibnamefont
  {Hager}}, \bibinfo {author} {\bibfnamefont {J.~G.}\ \bibnamefont {McNally}},
  \ and\ \bibinfo {author} {\bibfnamefont {T.}~\bibnamefont {Misteli}},\ }\href
  {\doibase 10.1016/j.molcel.2009.09.005} {\bibfield  {journal} {\bibinfo
  {journal} {Mol. Cell}\ }\textbf {\bibinfo {volume} {35}},\ \bibinfo {pages}
  {741} (\bibinfo {year} {2009})}\BibitemShut {NoStop}%
\bibitem [{\citenamefont {Vukojevic}\ \emph {et~al.}(2010)\citenamefont
  {Vukojevic}, \citenamefont {Papadopoulos}, \citenamefont {Terenius},
  \citenamefont {Gehring},\ and\ \citenamefont {Rigler}}]{vukojevic_2010}%
  \BibitemOpen
  \bibfield  {author} {\bibinfo {author} {\bibfnamefont {V.}~\bibnamefont
  {Vukojevic}}, \bibinfo {author} {\bibfnamefont {D.~K.}\ \bibnamefont
  {Papadopoulos}}, \bibinfo {author} {\bibfnamefont {L.}~\bibnamefont
  {Terenius}}, \bibinfo {author} {\bibfnamefont {W.~J.}\ \bibnamefont
  {Gehring}}, \ and\ \bibinfo {author} {\bibfnamefont {R.}~\bibnamefont
  {Rigler}},\ }\href {\doibase 10.1073/pnas.0914612107} {\bibfield  {journal}
  {\bibinfo  {journal} {Proc. Natl. Acad. Sci.}\ }\textbf {\bibinfo {volume}
  {107}},\ \bibinfo {pages} {4093} (\bibinfo {year} {2010})}\BibitemShut
  {NoStop}%
\bibitem [{\citenamefont {Chen}\ \emph {et~al.}(2014)\citenamefont {Chen},
  \citenamefont {Zhang}, \citenamefont {Li}, \citenamefont {Chen},
  \citenamefont {Revyakin}, \citenamefont {Hajj}, \citenamefont {Legant},
  \citenamefont {Dahan}, \citenamefont {Lionnet}, \citenamefont {Betzig},
  \citenamefont {Tjian},\ and\ \citenamefont {Liu}}]{chen_2014}%
  \BibitemOpen
  \bibfield  {author} {\bibinfo {author} {\bibfnamefont {J.}~\bibnamefont
  {Chen}}, \bibinfo {author} {\bibfnamefont {Z.}~\bibnamefont {Zhang}},
  \bibinfo {author} {\bibfnamefont {L.}~\bibnamefont {Li}}, \bibinfo {author}
  {\bibfnamefont {B.-C.}\ \bibnamefont {Chen}}, \bibinfo {author}
  {\bibfnamefont {A.}~\bibnamefont {Revyakin}}, \bibinfo {author}
  {\bibfnamefont {B.}~\bibnamefont {Hajj}}, \bibinfo {author} {\bibfnamefont
  {W.}~\bibnamefont {Legant}}, \bibinfo {author} {\bibfnamefont
  {M.}~\bibnamefont {Dahan}}, \bibinfo {author} {\bibfnamefont
  {T.}~\bibnamefont {Lionnet}}, \bibinfo {author} {\bibfnamefont
  {E.}~\bibnamefont {Betzig}}, \bibinfo {author} {\bibfnamefont
  {R.}~\bibnamefont {Tjian}}, \ and\ \bibinfo {author} {\bibfnamefont
  {Z.}~\bibnamefont {Liu}},\ }\href {\doibase 10.1016/j.cell.2014.01.062}
  {\bibfield  {journal} {\bibinfo  {journal} {Cell}\ }\textbf {\bibinfo
  {volume} {156}},\ \bibinfo {pages} {1274} (\bibinfo {year}
  {2014})}\BibitemShut {NoStop}%
\bibitem [{\citenamefont {Pedraza}\ and\ \citenamefont
  {Paulsson}(2008)}]{pedraza_2008}%
  \BibitemOpen
  \bibfield  {author} {\bibinfo {author} {\bibfnamefont {J.~M.}\ \bibnamefont
  {Pedraza}}\ and\ \bibinfo {author} {\bibfnamefont {J.}~\bibnamefont
  {Paulsson}},\ }\href {\doibase 10.1126/science.1144331} {\bibfield  {journal}
  {\bibinfo  {journal} {Science}\ }\textbf {\bibinfo {volume} {319}},\ \bibinfo
  {pages} {339} (\bibinfo {year} {2008})}\BibitemShut {NoStop}%
\bibitem [{\citenamefont {Maerkl}\ and\ \citenamefont
  {Quake}(2007)}]{maerkl_2007}%
  \BibitemOpen
  \bibfield  {author} {\bibinfo {author} {\bibfnamefont {S.~J.}\ \bibnamefont
  {Maerkl}}\ and\ \bibinfo {author} {\bibfnamefont {S.~R.}\ \bibnamefont
  {Quake}},\ }\href {\doibase 10.1126/science.1131007} {\bibfield  {journal}
  {\bibinfo  {journal} {Science}\ }\textbf {\bibinfo {volume} {315}},\ \bibinfo
  {pages} {233} (\bibinfo {year} {2007})}\BibitemShut {NoStop}%
\end{thebibliography}%

\end{document}